\documentclass[reprint, double column, superscriptaddress,prl, showkeys]{revtex4-1}

\usepackage{graphicx,epsfig}
\usepackage{amssymb}
\usepackage{amsmath}
\usepackage{bm}
\usepackage{textcomp}
\usepackage{color}
\usepackage{appendix}
\usepackage{soul}

\begin{document}
\setcounter{page}{1}

\title[]{Thermal and quantum melting phase diagrams for a magnetic-field-induced Wigner solid}
\author{Meng K. \surname{Ma}}
\author{K. A. \surname{Villegas Rosales}}
\author{H. \surname{Deng}}
\author{Y. J. \surname{Chung}}
\author{L. N. \surname{Pfeiffer}}
\author{K. W. \surname{West}}
\author{K. W. \surname{Baldwin}}
\affiliation{Department of Electrical Engineering, Princeton University, Princeton, New Jersey 08544, USA}
\author{R. \surname{Winkler}}
\affiliation{Department of Physics, Northern Illinois University, DeKalb, Illinois 60115, USA}
\author{M. \surname{Shayegan}}
\affiliation{Department of Electrical Engineering, Princeton University, Princeton, New Jersey 08544, USA}
\date{\today}

\begin{abstract}

A sufficiently large perpendicular magnetic field quenches the kinetic (Fermi) energy of an interacting two-dimensional (2D) system of fermions, making them susceptible to the formation of a Wigner solid (WS) phase in which the charged carriers organize themselves in a periodic array in order to minimize their Coulomb repulsion energy. In low-disorder 2D electron systems confined to modulation-doped GaAs heterostructures, signatures of a magnetic-field-induced WS appear at low temperatures and very small Landau level filling factors ($\nu\simeq1/5$). In dilute GaAs 2D \textit{hole} systems, on the other hand, thanks to the larger hole effective mass and the ensuing Landau level mixing, the WS forms at relatively higher fillings ($\nu\simeq1/3$). Here we report our measurements of the fundamental temperature vs. filling phase diagram for the 2D holes' WS-liquid \textit{thermal melting}. Moreover, via changing the 2D hole density, we also probe their Landau level mixing vs. filling WS-liquid \textit{quantum melting} phase diagram. We find our data to be in good agreement with the results of very recent calculations, although intriguing subtleties remain.

\end{abstract}

\maketitle  

The Wigner solid (WS), an ordered array of electrons, favored when the Coulomb repulsion energy dominates over the thermal and Fermi energies, is one of the longest-anticipated and most exotic correlated phases of a strongly-interacting electron system \cite{Wigner.PR.1934}.  In a low-disorder, two-dimensional electron system (2DES) under a large perpendicular magnetic field ($B$), the Fermi energy is quenched and the electrons condense into the lowest Landau level (LL). If the separation between the LLs is large compared to the Coulomb energy so that LL mixing (LLM) can be ignored, a magnetic-field-induced, 2D quantum WS is expected at very small LL filling factors ($\nu\lesssim 1/5$) \cite{Lozovik.JETP.Lett.1975, Lam.PRB.1984, Levesque.PRB.1984,Archer.PRL.2013}. There is, however, a close competition with interacting liquid phases, such as the fractional quantum Hall states (FQHSs) \cite{Tsui.PRL.1982}. In very high mobility 2DESs confined to GaAs quantum wells where LLM is small, insulating phases are seen near the FQHS at $\nu=1/5$, and are generally believed to signal the formation of a WS, pinned by the small but ubiquitous disorder \cite{Andrei.PRL.1988, Willett.PRB.1988, Jiang.PRL.1990, Goldman.PRL.1990, Williams.PRL.1991,Li.PRL.1991, Jiang.PRB.1991,Paalanen.PRB.1992, Goldys.PRB.1992,Kukushkin.Euro.Phys.Lett.1993,Shayegan.WS.Review.1997, Ye.PRL.2002,Pan.PRL.2002, Chen.Nat.Phys.2006, Tiemann.Nat.Phys.2014, Deng.PRL.2016, Jang.Ashoori.Nat.Phys.2017, Deng.PRL.2019}. Many properties of these insulating phases support the pinned WS picture \cite{Shayegan.WS.Review.1997}; these include the non-linear current-voltage and noise characteristics \cite{Goldman.PRL.1990, Li.PRL.1991,Jiang.PRB.1991,Williams.PRL.1991}, microwave resonances \cite{Andrei.PRL.1988, Williams.PRL.1991, Chen.Nat.Phys.2006,Ye.PRL.2002}, photoluminescence  \cite{Goldys.PRB.1992, Kukushkin.Euro.Phys.Lett.1993}, nuclear magnetic resonance features \cite{Tiemann.Nat.Phys.2014}, tunneling resonances \cite{Jang.Ashoori.Nat.Phys.2017}, and screening characteristics \cite{Deng.PRL.2019}. There is also a recent experiment in a GaAs bilayer electron system with very imbalanced densities where one layer is near $\nu=1/2$ and contains composite fermions while the other layer is at very low fillings ($\nu\ll 1/5$) and hosts a WS \cite{Deng.PRL.2016}. The commensurability oscillations of the composite fermions induced by the periodic potential of the WS layer are used to directly probe the lattice constant of the WS. 

The 2D \textit{hole} systems (2DHSs) in low-disorder GaAs quantum wells provide a particularly exciting platform for studies of the quantum WS phases, both at $B=0$ \cite{Yoon.PRL.1999,Manfra.PRL.2007,Qiu.Gao.PRL.2012} and at high $B$ \cite{Santos.PRL.1992,Santos.PRB.1992,Bayot.EPL.1994,Li.PRL.1997,Li.PRB.2000,Csathy.PRL.2005,Pan.PRB.2005,Qiu.Gao.PRL.2012,Jang.Ashoori.Nat.Phys.2017,Jo.PRL.2018,Csathy.PRL.2004,Knighton.PRB.2018}. The effective mass for holes in GaAs is $m^*\simeq0.5$ (in units of the free electron mass) \cite{Zhu.SSC.2007}, much larger than $m^*\simeq0.067$ for GaAs 2D electrons, rendering the 2DHS effectively more dilute and therefore more interacting; note that the $r_s$ parameter, the inter-particle distance in units of the effective Bohr radius, scales with $m^*$. Signatures of a quantum WS at $B=0$ have indeed been reported in dilute GaAs 2DHSs with very large $r_s$ \cite{Yoon.PRL.1999, Manfra.PRL.2007, Qiu.Gao.PRL.2012}. At high $B$, the larger $m^*$ means that the LL separation is small so that there is a significant mixing of the higher LLs into the collective states of the 2D system. (For our samples reported here, the LLM parameter $\kappa$, defined as the ratio of the Coulomb to cyclotron energies ranges between $\sim 5$ and $16$.)  Such LLM generally weakens the FQHSs, whose stability relies on short-range correlations, and favors the ground states with long-range order, such as the WS \cite{Yoshioka.JPSJ.1984, Yoshioka.JPSJ.1986, Zhu.PRL.1993, Price.PRL.1993, Platzman.PRL.1993, Ortiz.PRL.1993, Zhao.PRL.2018}. Consistent with this expectation, experiments on GaAs 2DHSs have indeed shown that the onset of the magnetic-field-induced WS moves to higher fillings ($\nu\simeq1/3$, compared to $\nu\simeq1/5$ for 2D electrons) \cite{Santos.PRL.1992, Santos.PRB.1992,Bayot.EPL.1994,Li.PRL.1997,Shayegan.WS.Review.1997,Li.PRB.2000,Pan.PRB.2005,Csathy.PRL.2005,Jo.PRL.2018,Csathy.PRL.2004,Knighton.PRB.2018}. A recent study on ZnO 2DESs with parameters similar to GaAs 2DHSs also shows the onset of the WS at $\nu\simeq1/3$ \cite{Maryenko.Nat.Comm.2018}. Here we present experiments on very low disorder 2DHSs confined to modulation-doped GaAs quantum wells, and probe two fundamental WS-liquid phase diagrams: a temperature vs. $\nu$ phase diagram for the thermal melting of the WS, and a $\kappa$ vs. $\nu$ diagram for its quantum melting. 

\begin{figure}[b!]
  \begin{center}
    \psfig{file=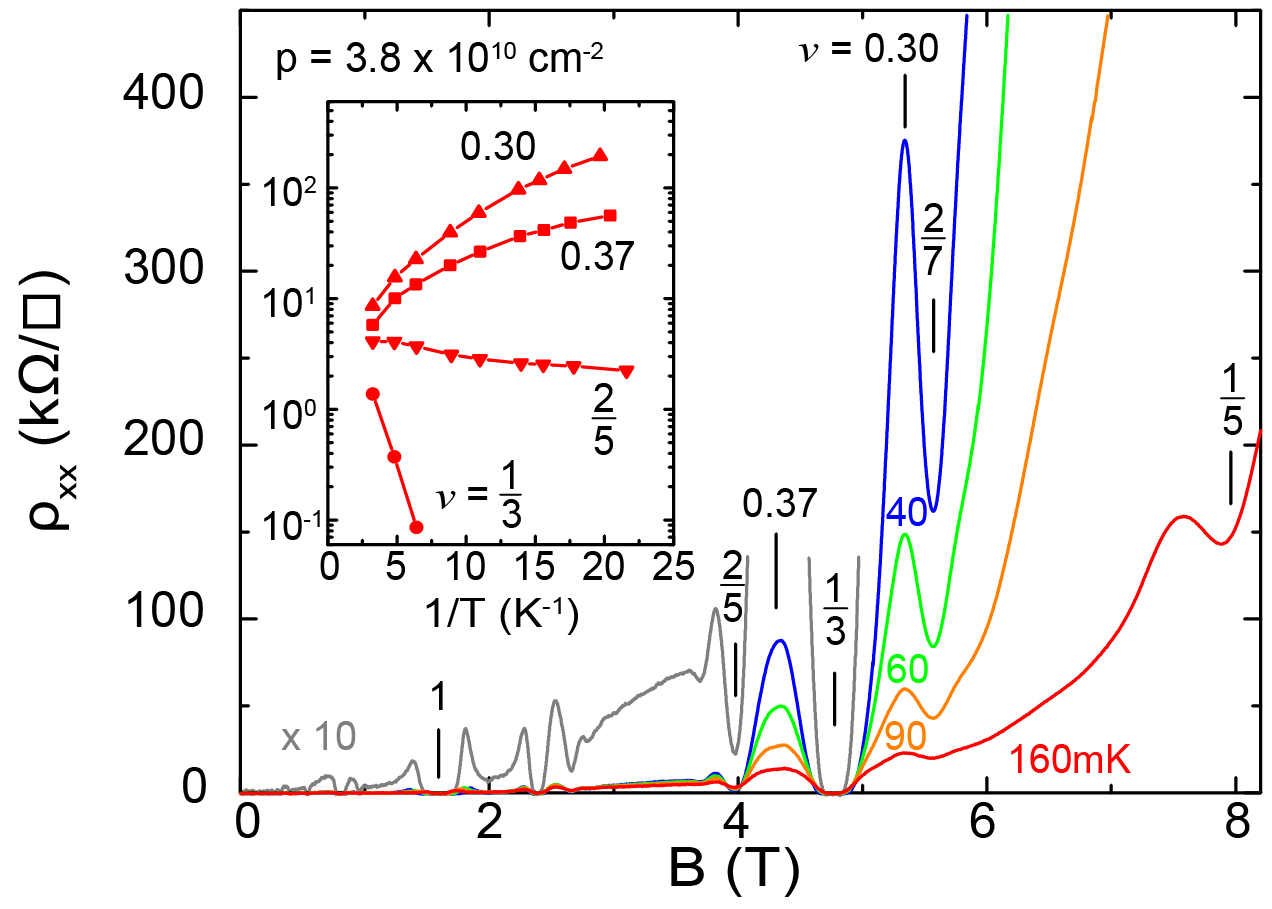, width=0.45
    \textwidth}
  \end{center}
  \caption{\label{transport} 
  Temperature dependence of the longitudinal resistivity $\rho_{xx}$ vs. magnetic field $B$ at $p=3.8$. The $y$-scale for the grey trace is expanded by a factor of $10$. The inset shows the Arrhenius plots of $\rho_{xx}$ at $\nu=0.30$, $0.37$, $2/5$ and $1/3$.
  }
  \label{fig:transport}
\end{figure}

We studied 2DHSs confined to modulation-doped, 30-nm-wide GaAs quantum wells (QWs) grown on GaAs (100) substrates. The details of the sample parameters are provided in the Supplemental Material (SM) \cite{supplemental}. The samples have 2DHS densities ($p$) ranging from $2.0$ to $7.9$, in units of $10^{10}$ cm$^{-2}$ which we will use throughout this paper, and their low-temperature mobility is $\simeq1.5 \times 10^{6}$ cm$^{2}$/Vs. We present data in the main text for two samples with densities $p=3.8$ and $7.9$; data for other densities are shown in the SM \cite{supplemental}. We performed all our measurements on 4 mm $\times$ 4 mm van der Pauw geometry samples, which are fitted with gate electrodes deposited on their top and bottom surfaces. The density in a given sample is tuned using both the front and back gates while keeping the charge distribution in the QW symmetric. We made measurements primarily in a dilution refrigerator with a base temperature of $\simeq40$ mK.

Figure \ref{fig:transport} shows the temperature dependence of longitudinal resistivity $\rho_{xx}$ vs. $B$ at $p=3.8$. The expanded (grey) trace at $\simeq40$ mK shows a series of FQHSs attesting to the good quality of the sample. At the highest temperature, there is even a hint of a  developing $\nu=1/5$ FQHS. The $\nu=1/3$ FQHS is fully developed and has a vanishing $\rho_{xx}$ minimum at the lowest temperatures. On the other hand, on its flanks (e.g., at $\nu=0.30$ and $0.37$), $\rho_{xx}$ has very high values, which decrease rapidly as temperature is raised. This insulating behavior is generally believed to signal a disorder-pinned WS state \cite{Santos.PRL.1992, Santos.PRB.1992,Li.PRL.1997,Shayegan.WS.Review.1997,Li.PRB.2000,Pan.PRB.2005,Csathy.PRL.2005,Jo.PRL.2018,Csathy.PRL.2004,Knighton.PRB.2018}, and can be seen more conveniently in the Arrhenius plot shown in Fig. \ref{fig:transport} inset. Also shown in this inset are the temperature dependence of $\rho_{xx}$ at $\nu=2/5$ and $1/3$. In contrast to the insulating behavior at $\nu=0.30$ and $0.37$, $\rho_{xx}$ at $\nu=1/3$ and $2/5$ decreases as temperature is lowered, and is activated at $\nu=1/3$ with an energy gap of $\simeq 1.76$ K. 

\begin{figure}[t!]
  \begin{center}
    \psfig{file=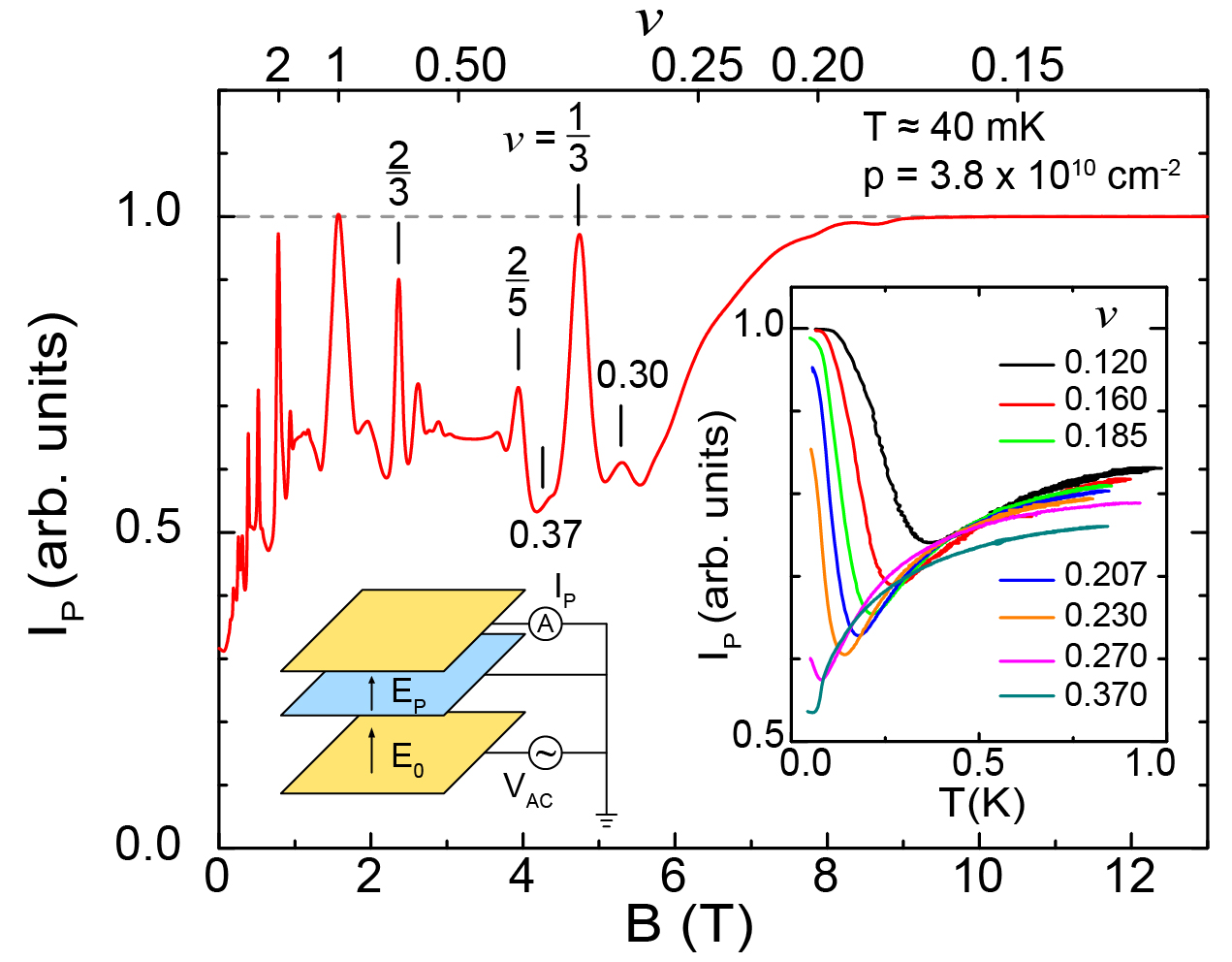, width=0.45\textwidth}
  \end{center}
  \caption{\label{Ip} 
Left inset: schematic of the measurement setup. Top and bottom yellow plates represent front and back gates and the middle blue plate the 2DHS layer. AC excitation voltage $V_{AC}$ is applied to the bottom gate which generates an electric field $E_0$, and subsequently a penetrating electric field $E_P$ as a function of the screening efficiency of the 2DHS. A current $I_P$ in response to $E_P$ is then measured. Trace in the main figure shows $I_P$, normalized to its maximum value, vs. $B$. Horizontal line marks the maximum of $I_P$ when the 2DHS screening is minimum. Right inset: $I_P$ vs. temperature traces for $\nu$ ranging from $0.120$ to $0.370$.
}
  \label{fig:Ip}
\end{figure}

We probe the thermal melting of the WS by monitoring the screening efficiency \cite{Deng.PRL.2019,Eisenstein.PRL.1992,Eisenstein.PRB.1994,Young.Nat.Phys.2018} of the 2DHS. This technique was used recently \cite{Deng.PRL.2019} to study the magnetic-field-induced WS in GaAs 2DESs near $\nu\simeq1/5$, and the deduced melting phase diagram was found to be in good agreement with previous measurements. The measurement setup is shown schematically in Fig. \ref{fig:Ip} inset. The top and bottom yellow plates represent the front and back gates. The blue layer in the middle represents the 2DHS we are probing. We apply an AC excitation voltage $V_{AC}$ of $1$ mV between the back and front gates at $22$ kHz as shown in the inset. This AC voltage generates an electric field $E_P$ penetrating through the 2DHS. The magnitude of $E_P$ depends on the screening efficiency of the 2DHS. The magnitude of the penetrating current $I_P$ is then probed in response to $E_P$.

The trace in Fig. \ref{fig:Ip} shows $I_P$ vs. $B$ at our base temperature ($\simeq40$ mK) and $p=3.8$. At fillings where the 2DHS is in an integer or FQHS, its bulk is incompressible and the screening is minimal. As a result, $I_P$ shows a maximum. When the bulk is compressible between the QHSs, $I_P$ comes down as a result of the increasing screening efficiency of the 2DHS. At $\nu=0.30$, where the WS phase develops, $I_P$ shows a local maximum, consistent with the WS phase being insulating and having, therefore, lower screening efficiency. $I_P$ at $\nu=0.37$ shows a ``shoulder" at this density, but develops into a well-defined local maximum at lower densities \cite{supplemental}. At very high $B$, beyond $\simeq8$ T, the 2DHS becomes strongly insulating and $I_P$ approaches the same value it has at the strongest QHSs, consistent with the screening efficiency being minimal. 

The right inset in Fig. \ref{fig:Ip} shows the temperature dependence of $I_P$ at different $\nu$. At $\nu=0.120$, $I_P$ starts with high value at the lowest temperature, consistent with an insulating WS. At the highest temperatures, where we expect the WS to have melted, $I_P$ saturates at a value which is lower than its maximum value. This is consistent with a compressible liquid phase which has a higher screening efficiency than the WS. However, as temperature is raised, instead of decreasing monotonically from its low-temperature value and saturating at the high-temperature limit, $I_P$ shows a well-defined minimum at a critical temperature $T_C$. This temperature dependence is generic for all the traces shown in Fig. \ref{fig:transport} inset except for $\nu=0.370$ and $\nu=0.270$, where $I_P$ at the lowest temperature is lower than its high-temperature limit. This is because the lowest temperature achieved in our measurements ($T\simeq 40$ mK) is close to $T_C$ for these two fillings; we expect $I_P$ to increase if lower temperatures were accessible. 

The data shown in Fig. \ref{fig:Ip} inset suggest that the 2DHS becomes particularly efficient at screening near $T_C$. A qualitatively similar behavior was recently seen in low-density GaAs 2D $electron$ systems \cite{Deng.PRL.2019}. Associating $T_C$ with the melting temperature of the WS, Ref. \cite{Deng.PRL.2019} found the measured dependence of $T_C$ on $\nu$ to be consistent with the WS melting phase diagrams reported previously for the magnetic-field-induced WS in GaAs 2DESs. It is not clear why a WS should become particularly efficient at screening as it melts. It is possible that the minimum in $I_P$ signals the presence of an intermediate phase near the melting temperature, as has been suggested in a recent report \cite{Knighton.PRB.2018}. Alternatively, very recent calculations \cite{Delacretaz.PRB.2019} suggest that dissipation from mobile dislocations and uncondensed charge carriers become especially important near the melting of the WS phase. It is possible that they contribute to the extra screening at the melting. 

\begin{figure}[t!]
  \begin{center}
    \psfig{file=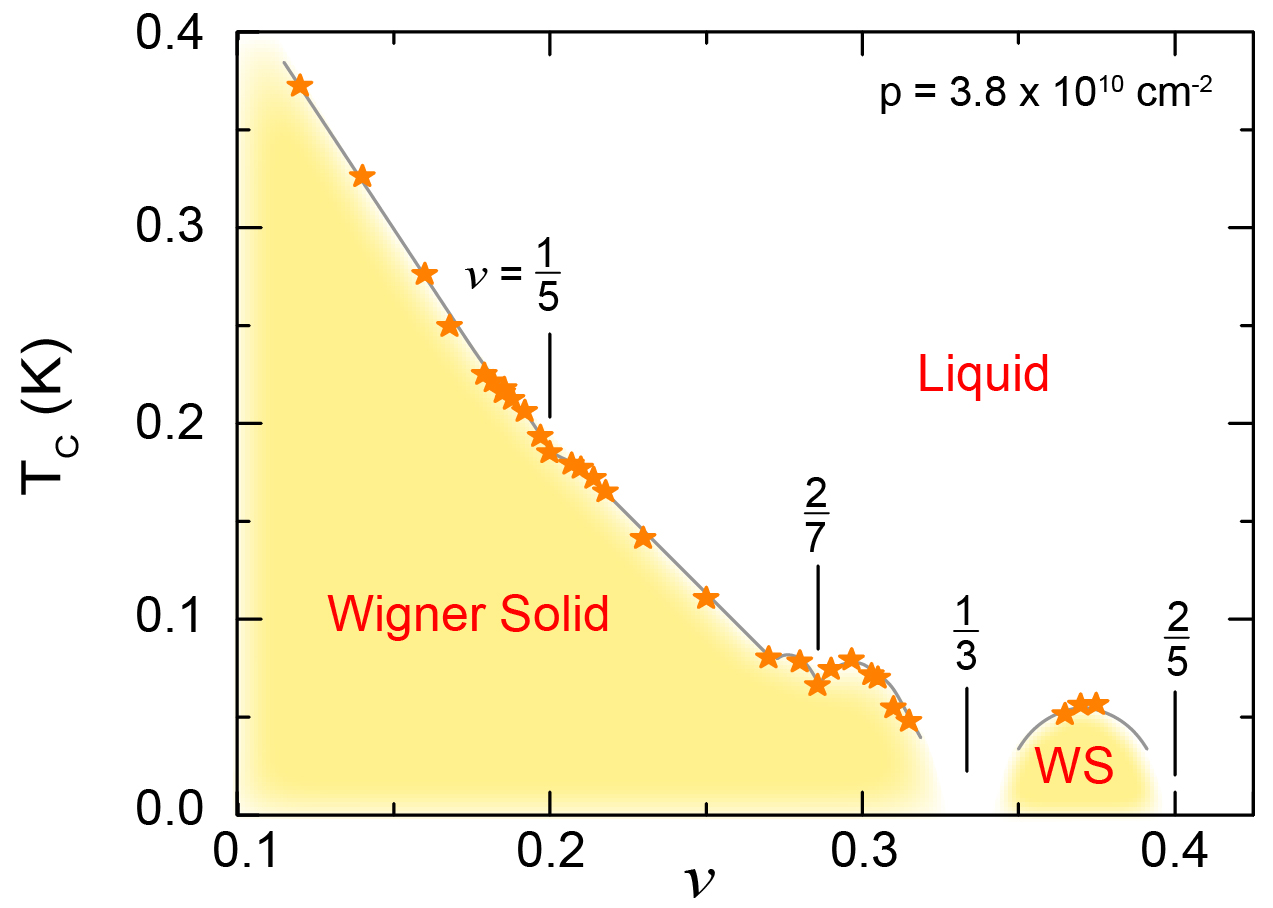, width=0.47\textwidth}
  \end{center}
  \caption{\label{phase1} 
  WS \textit{thermal melting} phase diagram. The yellow and white regions indicate the solid and liquid phases, respectively. The grey line connecting the measured data points is a guide to the eye.
  }
  \label{fig:phase1}
\end{figure}

Associating $T_C$ with the melting temperature of the WS, a plot of our measured $T_C$ vs. $\nu$, as shown in Fig. \ref{fig:phase1}, provides the WS \textit{thermal melting} phase diagram of a 2DHS at $p=3.8$. As $\nu$ increases from small values, $T_C$ decreases until the WS phase is ``interrupted" by the well-developed $\nu=1/3$ FQHS. When $\nu$ is higher than $1/3$, there is a reentrant WS phase between the $1/3$ and $2/5$ FQHSs, around $\nu\simeq0.37$. We note that our $T_C \simeq50$ mK at $\nu=0.37$ is consistent with the WS melting temperature reported in Ref. \cite{Knighton.PRB.2018} for a 2DHS with a similar density at $\nu=0.375$.   

\begin{figure}[t!]
  \begin{center}
    \psfig{file=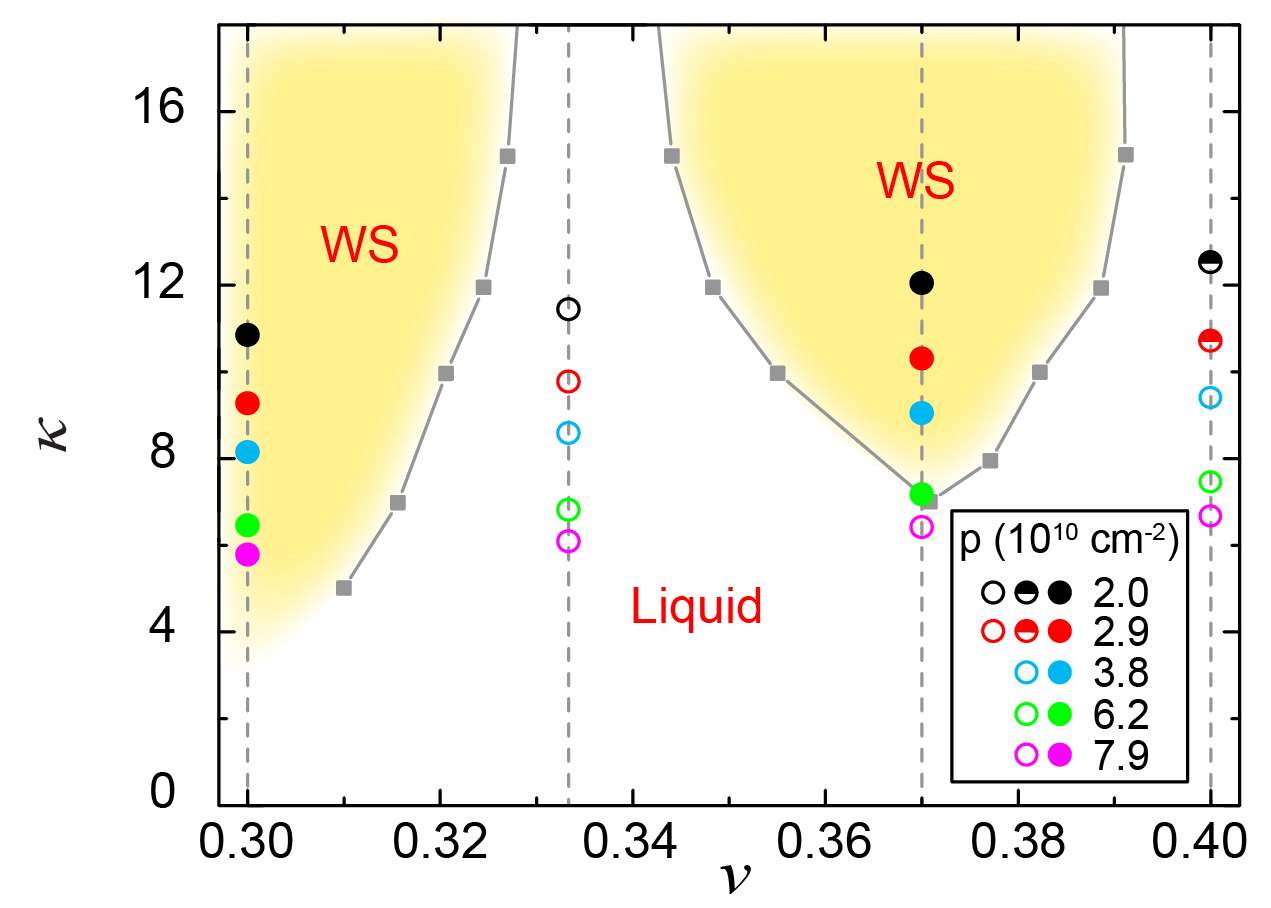, width=0.47\textwidth}
  \end{center}
  \caption{\label{phase2} 
  WS \textit{quantum melting} phase diagram, with the Landau level mixing parameter $\kappa$ and $\nu$ for the axes. The grey solid squares connected by the guide-to-the-eye lines are from theoretical calculations \cite{Zhao.PRL.2018}. The yellow and white regions indicate the predicted WS and liquid phases. The color-coded circles represent experimental data points deduced from measurements at six different densities, as listed in the inset box. The closed and open circles indicate WS and liquid phases, respectively. The half-filled circles are used to imply a close competition between the WS and the liquid phase.
  }
  \label{fig:phase2}
\end{figure}

The competition between the WS and FQHS liquid phases depends on the mixing between the LLs \cite{Santos.PRL.1992, Santos.PRB.1992,Li.PRL.1997,Li.PRB.2000,Csathy.PRL.2005,Pan.PRB.2005,Jo.PRL.2018,Yoshioka.JPSJ.1984,Yoshioka.JPSJ.1986, Zhu.PRL.1993,Price.PRL.1993,Platzman.PRL.1993,Ortiz.PRL.1993,Zhao.PRL.2018,Csathy.PRL.2004,Knighton.PRB.2018}. This is often quantified in terms of the LLM parameter $\kappa$, defined as the ratio between the Coulomb energy and the LL separation: $\kappa=(e^2/4 \pi \epsilon_0 \epsilon l_B)/(\hbar eB/m^*)$, where $l_B=\sqrt{\hbar /eB}$ is the magnetic length. Note that $\kappa\propto m^*$. When $\kappa$ is large, the mixing with the higher LLs reduces the FQHS energy gaps and favors the formation of a WS at filling factors higher than $1/5$ \cite{Santos.PRL.1992, Santos.PRB.1992,Li.PRL.1997,Li.PRB.2000,Csathy.PRL.2005,Pan.PRB.2005,Jo.PRL.2018,Yoshioka.JPSJ.1984,Yoshioka.JPSJ.1986, Zhu.PRL.1993,Price.PRL.1993,Platzman.PRL.1993,Ortiz.PRL.1993,Zhao.PRL.2018,Csathy.PRL.2004,Knighton.PRB.2018}. Recent theoretical work by Zhao \textit{et al.} \cite{Zhao.PRL.2018} directly mapped out a zero-temperature phase diagram for the quantum melting of the WS in the $\kappa$-$\nu$ space. The calculated phase diagram is reproduced in Fig. \ref{fig:phase2} for a direct comparison with our experimental results.

Our GaAs 2DHSs allow us to test the role of LLM. Compared to the GaAs 2DES, the 2DHSs have large $m^*$. Because of the non-parabolicity of the valence bands and spin-orbit coupling, however, $m^*$ for 2DHS is intrinsically complex and depends on the specific sample parameters such as the QW width and the symmetry of the charge distribution \cite{Winkler.Book.2003}. For a systematic study, it is therefore essential to use both the front and back gates to keep the hole charge distribution in the QW symmetric while changing the density. Cyclotron resonance experiments \cite{Zhu.SSC.2007} on 2DHSs confined to symmetric, 30-nm-wide QWs grown on GaAs (100) substrates yield a weakly density-dependent $m^*\simeq0.48$ in the density range we studied here. We use this value of $m^*$ to determine values of $\kappa$ at a set of representative filling factors $\nu=0.30$, $1/3$, $0.37$ and $2/5$ for our samples, and show these in Fig. \ref{fig:phase2} using symbols described in the inset.

\begin{figure}[t!]
  \begin{center}
    \psfig{file=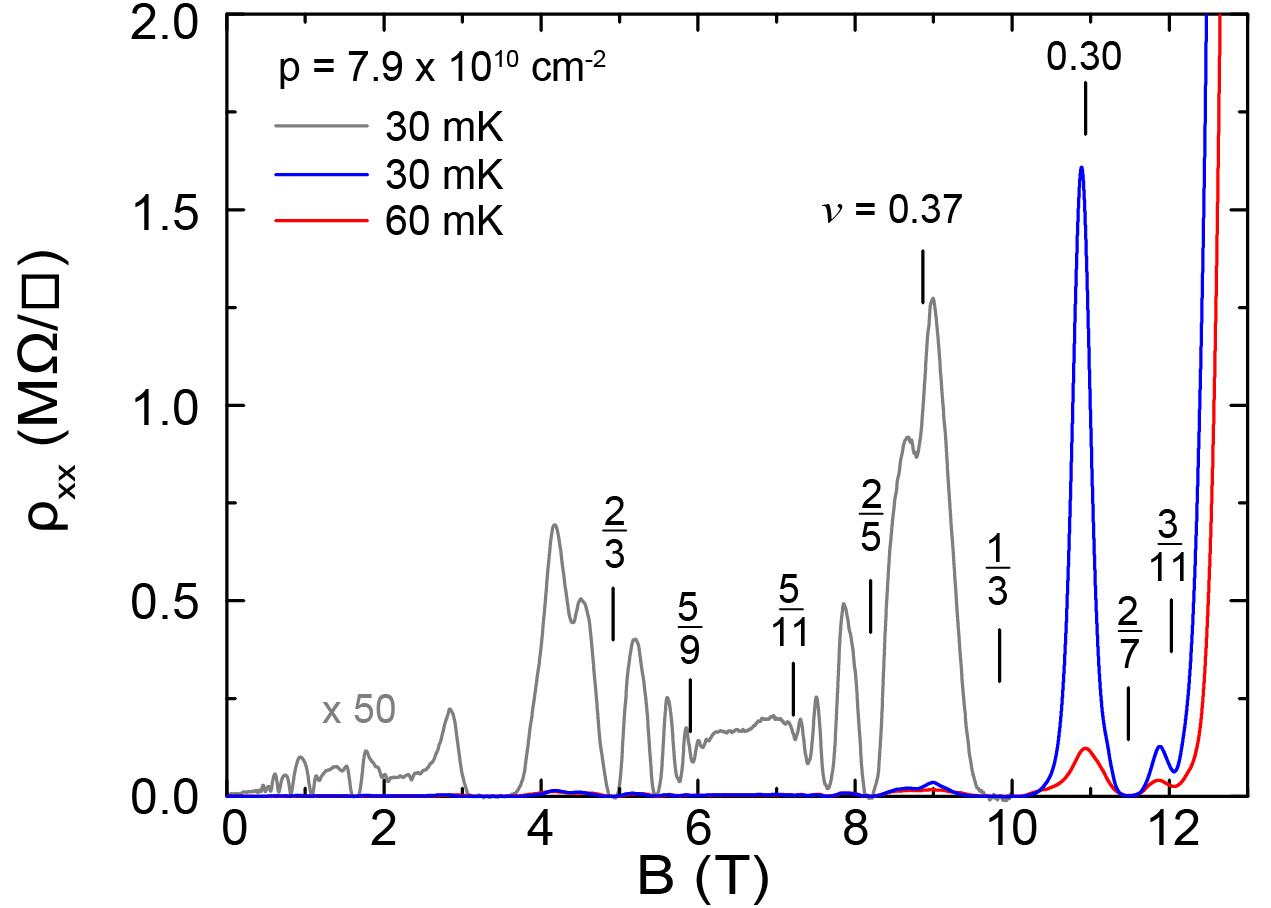, width=0.47\textwidth}
  \end{center}
  \caption{\label{transport_2} 
 Magneto-resistivity data for a 2DHS with $p=7.9$ at $30$ and $60$ mK. The $y$-scale for the grey trace is expanded by factor of $50$ to show the numerous observed FQHSs, attesting to the high quality of the 2DHS.
  }
  \label{fig:transport_2}
\end{figure}

For $p=3.8$, the experimental data are represented by blue circles in Fig. \ref{fig:phase2}. Data at all four fillings are consistent with the calculation results: as $\nu$ decreases, the 2DHS ground state changes from a FQHS at $\nu=2/5$ to a WS at $0.37$, then to a FQHS at $1/3$, and finally back to a WS at $0.30$. In order to lower $\kappa$, we made measurements on a higher density 2DHS. The $\rho_{xx}$ vs. $B$ data for this sample are shown in Fig. \ref{fig:transport_2}. At this density, well-developed FQHSs are seen at $\nu=1/3$, $2/5$, and $2/7$. Moreover, in contrast to the trace at $p=3.8$ (Fig. \ref{fig:transport}), $\rho_{xx}$ at $\nu=0.37$ has comparable value to $\rho_{xx}$ at higher fillings, and depends only very weekly on temperature. This implies that the ground state at $\nu=0.37$ is not a WS at $p=7.9$. On the other hand, similar to the data for $p=3.8$, the trace in Fig. \ref{fig:transport_2} shows a very large and strongly temperature dependent $\rho_{xx}$ peak at $\nu=0.30$, consistent with a pinned WS. We show the four experimental points for $p=7.9$ at $\nu=0.30$, $1/3$, $0.37$, and $2/5$, in Fig. \ref{fig:phase2} by purple circles. The data are again consistent with the theoretical phase diagram: as $\kappa$ is reduced, the WS phase at $\nu=0.37$ disappears but it is still present at $\nu=0.30$.

We also performed measurements at three other densities, $p=6.2$, $2.9$, and $2.0$; the results are presented in the SM \cite{supplemental}, and are summarized in Fig. \ref{fig:phase2}. For $p=6.2$, the results are consistent with the theoretical phase diagram. For the lowest two densities, $p=2.9$ and $2.0$, however, there is a hint of a FQHS at $\nu=2/5$, but the data suggest a competition with an insulating phase, signaled by a rise in $\rho_{xx}$ as the temperature is lowered. This might indicate an apparent discrepancy between the experimental data and the theoretical phase diagram, which predicts that the ground state should be a FQHS (liquid) phase at $\nu=2/5$ in the entire range of $\kappa$ in Fig. \ref{fig:phase2}. We believe that disorder, whose role certainly increases at very low densities but is neglected in theory of Ref. \cite{Zhao.PRL.2018}, is at least partly responsible for the discrepancy \cite{supplemental}. It is worth remembering that, in early studies of GaAs 2DES, qualitatively similar observations were made. Early samples, which had lower quality, showed a competition between a FQHS and an insulating phase at $\nu=1/5$ \cite{Willett.PRB.1988}, and a clear FQHS with a vanishing $\rho_{xx}$ at the lowest temperature was only seen when samples of much better quality were available \cite{Jiang.PRL.1990}.

In conclusion, we report a thermal melting phase diagram for the magnetic-field-induced WS in GaAs 2DHSs deduced from its screening efficiency. The phase diagram shows the clear reentrant behavior of the WS around the FQHS at $\nu=1/3$, and provides data for a quantitative comparison with future theoretical calculations. We also systematically study the quantum melting of the WS as a function of LLM, varied by changing the 2DHS density. While we find good overall agreement with the results of calculations, we would like to emphasize the complexity of the 2DHS LL diagram \cite{Winkler.Book.2003}. As discussed in more detail in the SM \cite{supplemental}, the 2DHS LLs are non-linear and also can cross as a function of magnetic field. Moreover, the interaction between holes is subtle because of the multi-component and mixed (spin and orbital) nature of the hole states. These make a quantitative assessment of the role of LLM challenging. We hope that our experimental data provide incentive for a more precise theoretical evaluation of the role of LLM, as well as disorder, in the competition between the WS and FQHS phases in GaAs 2DHSs.

\begin{acknowledgments}

We acknowledge support by the DOE BES (No. DE-FG02-00-ER45841) Grant for measurements, and the NSF (Grants No. DMR 1709076, No. MRSEC DMR 1420541, and No. ECCS 1906253), and the Gordon and Betty Moore Foundation’s EPiQS Initiative (Grant No. GBMF9615) for sample fabrication and characterization. M.S. also acknowledges a QuantEmX travel grant from Institute for Complex Adaptive Matter and the Gordon and Betty Moore Foundation through Grant No. GBMF5305. We thank J.K. Jain and L.W. Engel for illuminating discussions.

\end{acknowledgments}

\clearpage

\appendix

\section{Supplemental Material}
In this Supplemental Material we present the details of the sample parameters and experimental setup. We also provide the results of self-consistent calculations for the subband energies, charge distribution, and Landau levels (LLs) in our two-dimensional hole systems (2DHSs), and discuss the implications of the strong non-linearity of the LLs. We also show additional measurement results, and discuss the role of disorder. 

\section{Sample parameters and experimental set up}

We studied 2DHSs in three samples. In all three samples, the 2DHS is confined to a $30$-nm-wide GaAs quantum well (QW) grown on a GaAs (100) substrate, and is flanked on each sides by Al$_{x}$Ga$_{1-x}$As spacer layers followed by carbon $\delta$-dopings. The structure is then buried under a $200$-nm-thick Al$_{x}$Ga$_{1-x}$As layer and finished with a $28$-nm-thick GaAs cap layer on top. The details of the sample parameters are listed in Table \ref{tab:table1}.

\begin{table}[tbh!]
\centering
\begin{tabular}{|c|c|c|c|c|}
\hline
Sample & 
\begin{tabular}[c]{@{}c@{}}$x$\\ ($\%$)\end{tabular} & \begin{tabular}[c]{@{}c@{}}$s$\\ (nm)\end{tabular} & \begin{tabular}[c]{@{}c@{}}$p$\\ ($10^{10}$ cm$^{-2}$)\end{tabular} & \begin{tabular}[c]{@{}c@{}}$\mu$ \\ ($10^6$ cm$^{2}$/Vs)\end{tabular} \\ \hline
A & 30 & 510 & 3.8 & 1.3 \\ \hline 
B & 30 & 278 & 7.9 & 1.8 \\ \hline 
C & 24 & 400 & 3.7 & 1.3 \\ \hline
\end{tabular}
\caption{Sample parameters. $x$ is the Al mole fraction in the barrier, $s$ is the spacer layer thickness, $p$ is the as-grown 2DHS density, and $\mu$ is the low-temperature ($T=0.3$ K) mobility.}
\label{tab:table1}
\end{table}

We performed all our measurements on 4 mm $\times$ 4 mm van der Pauw geometry samples, with InZn on the corners and sides annealed at 380 \textdegree C to make eight contacts to the 2DHS. For samples A and C, Ti/Au was deposited on top and used as front gates, and back gates were made by placing the samples on top of melted In. Both front and back gates are used to tune the density while keeping the charge distribution in the QW symmetric. We studied two densities $p=6.2$ and $3.8$ $\times 10^{10}$ cm$^{-2}$ for sample A, $p=7.9 \times 10^{10}$ cm$^{-2}$ for sample B, and three densities $p=3.7$, $2.9$, and $2.0 \times 10^{10}$ cm$^{-2}$ for sample C. The results for sample A at $p=3.8 \times 10^{10}$ cm$^{-2}$ and sample B at $p=7.9 \times 10^{10}$ cm$^{-2}$ are presented in the main text. Samples A and C were measured in a cryogen-free dilution refrigerator with a base temperature of $\simeq40$ mK, while sample B was measured in a wet dilution refrigerator with a base temperature of $\simeq30$ mK. All in-plane, magneto-transport measurements in this study were done using a low-frequency lock-in technique at $3$ to $7$ Hz frequency with excitation currents ranging from $1$ to $10$ nA. For the screening-efficiency (capacitance) measurements, we applied $1$ mV AC excitation voltage $V_{AC}$ to the back gate at various frequencies around $20$ kHz while keeping the 2DHS grounded, and measured the penetrating current from the front gate. Note that, because of the large distance between the back gate and the 2DES ($\simeq 500$ $\mu$m), the density modulation due to $V_{AC}$ is negligible.

\section{Subband energies and Landau levels}

\begin{figure*}[t!]
  \begin{center}
    \psfig{file=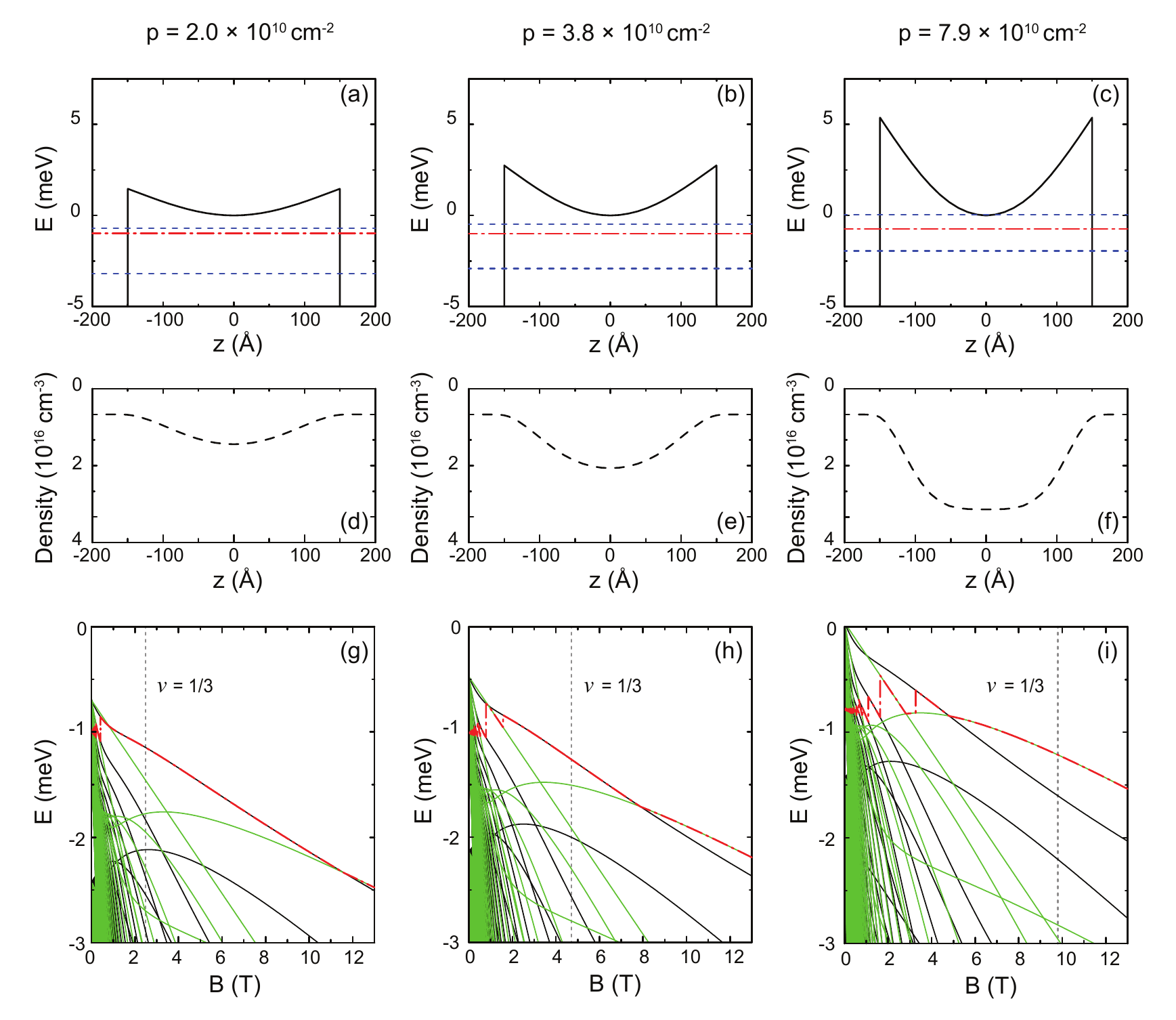, width=0.99\textwidth}
  \end{center}
  \caption{Panels (a)-(c) show the calculated, self-consistent, valence-band confinement potential (black curves) and the relevant subband and Fermi energies for 2DHSs confined in a $30$-nm-wide GaAs QW with densities $p=2.0$, $3.8$, and $7.9 \times 10^{10}$ cm$^{-2}$. The blue dashed lines in each figure show the energies of the first two subbands, and the red dash-dotted line the Fermi energy. Panels (d)-(f) show the corresponding self-consistent charge distributions in the QW. Panels (g)-(i) show the Landau level diagrams in the relevant magnetic field range. Black and green lines indicate Landau levels of opposite parity. The grey dashed lines mark the magnetic field positions of $\nu=1/3$ for different densities, and the red dash-dotted line traces the Fermi energy.}
  \label{fig:newfigS3}
\end{figure*}

\begin{figure}[b!]
  \begin{center}
    \psfig{file=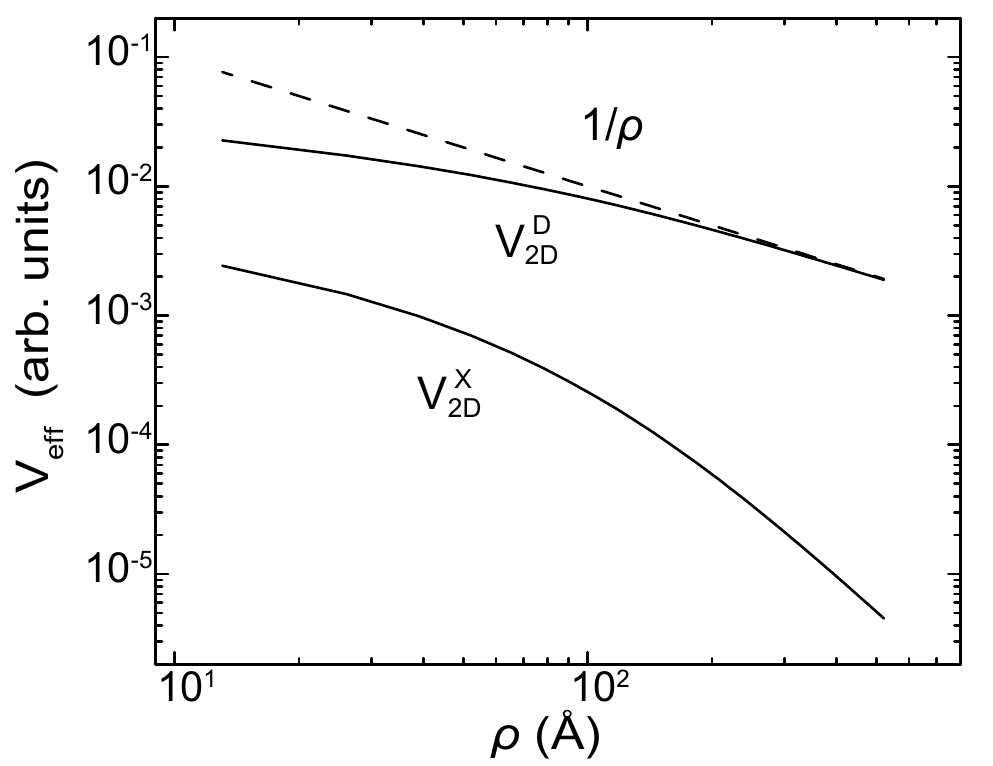, width=0.46\textwidth}
  \end{center}
  \caption{Effective 2D Coulomb interaction between the two uppermost Landau levels in Fig.~\ref{fig:newfigS3}(h) (density $p = 3.8 \times 10^{10}$~cm$^{-2}$) at $B = 8$~T. (Dashed lines) Bare 2D Coulomb interaction $1/\rho$. Similar results are obtained over the entire magnetic field range in Figs.~\ref{fig:newfigS3}(g)-(i) for any pair of Landau levels of opposite parity.}
  \label{fig:effCoul}
\end{figure}

We have performed self-consistent calculations of the hole subband structure using the multiband envelope function approximation based on the $8 \times 8$ Kane Hamiltonian \cite{Winkler.Book.2003} and using the Hartree approximation. Figure \ref{fig:newfigS3} shows the confinement potential, charge distribution and LL diagram for 2DHSs confined in a $30$-nm-wide GaAs QW with densities $p=2.0$, $3.8$, and $7.9 \times 10^{10}$ cm$^{-2}$. Figures \ref{fig:newfigS3}(a) to (c) indicate that, as the density increases, the separation between the first and the second subband energies decreases from $2.49$ meV at $p=2.0 \times 10^{10}$ cm$^{-2}$ to $1.99$ meV at $p=7.9 \times 10^{10}$ cm$^{-2}$. For all three densities, the Fermi energy, as indicated by the red dash-dotted line, lies well above the second subband, leading to a single-layer-like charge distribution as shown in Figs. \ref{fig:newfigS3}(d) to (f).

The calculated LL diagrams are shown in Figs. \ref{fig:newfigS3}(g) to (i). The LLs are non-linear and show multiple crossings as a function of magnetic field, which is very different from what one would expect based on a constant effective mass. Such complex LL diagrams make a quantitative assessment of the role of LLM very challenging. We note that in the analysis of the experimental data, we calculate the LLM parameter $\kappa$ based on the effective mass measured via cyclotron resonance at low magnetic fields \cite{Zhu.SSC.2007}. As described in the main text (e.g., see Fig. 4), we find good agreement between our experimental data and the theoretical quantum melting phase diagram calculated by Zhao $et$ $al.$ \cite{Zhao.PRL.2018}. Note that the parameter $\kappa$ used in the calculations of Ref. \cite{Zhao.PRL.2018} is based on a single value for the effective mass, and assuming LLs which have a simple, linear dependence on magnetic field. However, considering the LL diagrams shown in Fig. \ref{fig:newfigS3}, one should be cautious about the implications of such agreement. 

For all three densities in Fig. \ref{fig:newfigS3}, the calculations indicate a crossing of the two highest LLs at some fractional filling factor. Na\"{\i}vely, this corresponds to a vanishing cyclotron frequency (i.e., an infinite effective mass) and thus a diverging LLM parameter $\kappa$. However, we argue that due to the intricate nature of hole LLs, the LLM is greatly suppressed for this pair of LLs. To show this we consider the effective 2D interaction that is frequently used to approximate the Coulomb interaction between charge carriers in quasi-2D systems. In a one-band model appropriate for electrons, the effective 2D interaction reads \cite{Ando.RMP.1982, Jain.Book.2007}:
\begin{equation}
  \label{eq:coul-eff}
 V_\mathrm{2D}^\ast (\bm{\rho}) = \int dz_1 \int dz_2 \, 
 \frac{|\xi(z_1)|^2 \, |\xi(z_2)|^2}{\sqrt{\rho^2 + (z_1 - z_2)^2}} \; ,
\end{equation}
where $\bm{\rho} = (x,y)$ are the in-plane coordinates and $\xi(z)$ is the wave function for the out-of-plane motion. At short distances $\rho \lesssim w$ ($w$ is the width of the quasi-2D system), $V_\mathrm{2D}^\ast (\rho)$ is weaker than the bare 2D Coulomb interaction $V_\mathrm{2D} (\rho) = 1 / \rho$; and $V_\mathrm{2D}^\ast (\rho)$ approaches $V_\mathrm{2D} (\rho)$ for large distances $\rho \gg w$.

We want to generalize $V_\mathrm{2D}^\ast (\rho)$ to the case appropriate for holes where the charge carriers are characterized by multicomponent envelope functions \cite{Winkler.Book.2003}:
\begin{equation}\label{eq:EFAgen}
  \Psi_\alpha (\bm{\rho}, z) =
  \sum_i \phi_\alpha^i (\bm{\rho}) \:
  \xi_\alpha^i (z) \: u^i (\bm{\rho}, z) \; ,
\end{equation}
where $\phi_\alpha^i (\bm{\rho})$ and $\xi_\alpha^i (z)$ denote the in-plane and out-of-plane parts of the $i$th spinor component of $\Psi_\alpha$ in the basis of the bulk band-edge Bloch functions $u^i (\bm{\rho},z)$. The symbol $\alpha$ represents a generic index for the wave functions (\ref{eq:EFAgen}). In the present case $\alpha$ stands for the LL index. Ignoring the $i$-dependence of $\phi_\alpha^i (\bm{\rho})$, we obtain, similar to Eq.\ (\ref{eq:coul-eff}), the direct Coulomb interaction:
\begin{subequations}
  \label{eq:coul}
\begin{equation}
  \label{eq:coul:direct}
  V_\mathrm{2D}^\mathrm{D} (\rho) 
  = \sum_{i, j} \int \! dz_1 \int \! dz_2 \; \frac{
   |\xi_\alpha^i (z_1)|^2 \, |\xi_\beta^j (z_2)|^2}
   {\sqrt{\rho^2 + (z_1 - z_2)^2}}
 \end{equation}
and the exchange Coulomb interaction:
\begin{equation}
  \label{eq:coul:exchange}
  V_\mathrm{2D}^\mathrm{X} (\rho)
  = \sum_{i, j} \int \!\! dz_1 \! \int \!\! dz_2 \; \frac{
   \xi_\alpha^{j \ast} (z_1) \, \xi_\beta^{i \ast} (z_2) \,
   \xi_\alpha^i (z_2) \, \xi_\beta^j (z_1)}
   {\sqrt{\rho^2 + (z_1 - z_2)^2}}
   \; .
\end{equation}
\end{subequations}
In a one-band model, both $V_\mathrm{2D}^\mathrm{D} (\rho)$ and $V_\mathrm{2D}^\mathrm{X} (\rho)$ reduce to Eq.\ (\ref{eq:coul-eff}). We may expect that in a 2D jellium model with a homogenous background ensuring charge neutrality, the direct Coulomb term (\ref{eq:coul:direct}) can be ignored, i.e., the Coulomb interaction between the charge carriers is represented by the exchange term (\ref{eq:coul:exchange})
\cite{Kernreiter.PRB.2013}.

For the 2D hole systems studied here, spin-orbit coupling is large, so that spin is not a good quantum number for the states (\ref{eq:EFAgen}). However, for symmetric QWs as in the present experiments, the envelope functions $\Psi_\alpha (\bm{\rho}, z)$ characterizing the LLs are eigenstates of parity \cite{Andreani.PRB.1987}, as illustrated in Figs.~\ref{fig:newfigS3}(g)-(i), where black and green lines indicate LLs of opposite parity. In particular, it turns out that the envelope functions $\Psi_\alpha (\bm{\rho}, z)$ and $\Psi_\beta (\bm{\rho}, z)$ for the two lowest LLs have opposite parity, i.e., for each $i$ the spinors $\xi_\alpha^i (z)$ and $\xi_\beta^i (z)$ have opposite parity. The exchange interaction $V_\mathrm{2D}^\mathrm{X} (\rho)$ between these LLs is thus greatly reduced in magnitude, in particular for large distances $\rho$, when each $z$ integral is effectively an integral over an odd function (Fig.~\ref{fig:effCoul}). These theoretical considerations are consistent with our interpretation of the experiments using a large, but finite cyclotron effective mass, as cyclotron absorption is forbidden between states of opposite parity.

The suppression of the exchange interaction and thus the suppression of LLM between the two highest LLs is similar to the well-known fact that, if spin is a good quantum number, the exchange interaction acts only between charge carriers with the same spin orientation. It is also closely related to the fact that the enhancement of the Coulomb interaction in low-density quasi-2D hole systems can be greatly reduced in magnitude compared with the well-known enhancement of the Coulomb interaction in low-density quasi-2D electron systems, in particular when the quasi-2D hole system is spin-polarized \cite{Kernreiter.PRB.2013, Winkler.PRB.2005}. A more detailed theoretical study of these aspects will be published elsewhere. 

\section{Frequency dependence of the critical temperature and comparison with transport measurements}

\begin{figure*}[t!]
  \begin{center}
    \psfig{file=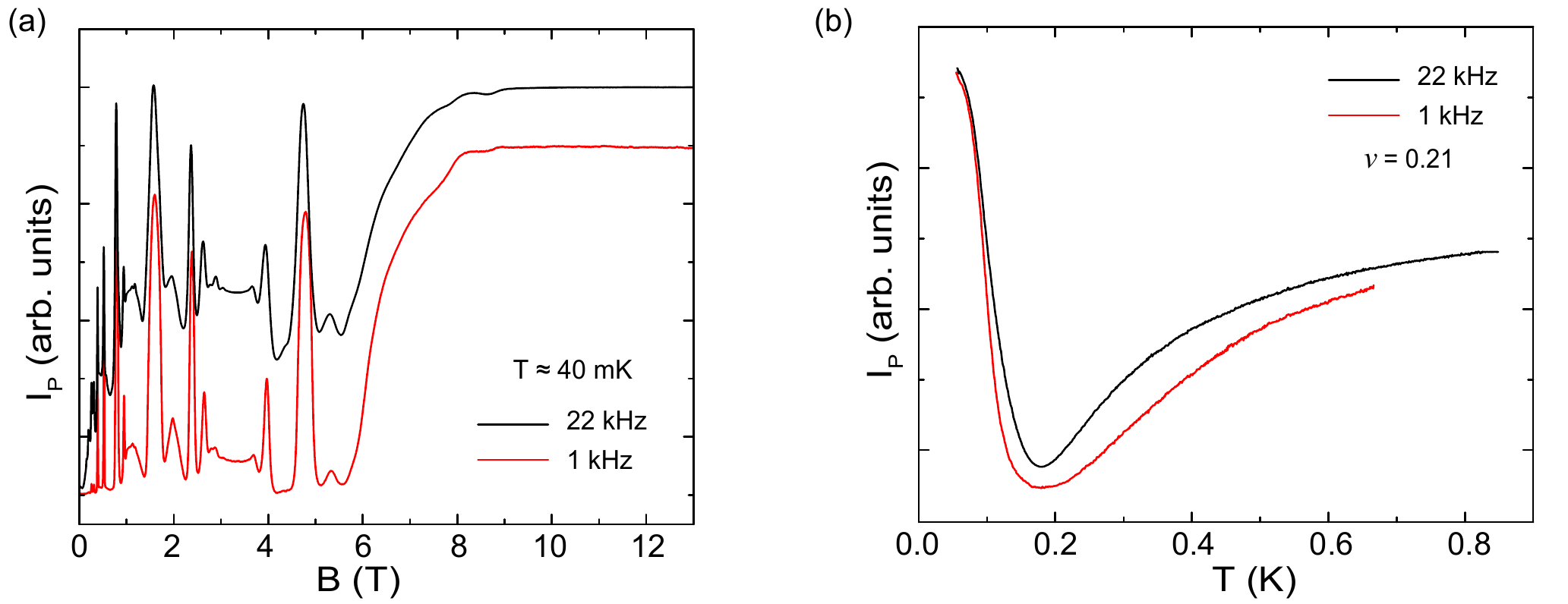, width=1\textwidth}
  \end{center}
  \caption{\label{newfigS1}
  Frequency dependence measurements for sample A at $p=3.8 \times 10^{10}$ cm$^{-2}$. (a) The penetrating current $I_P$ vs. magnetic field for measurement frequencies $22$ kHz and $1$ kHz. (b) $I_P$ vs. temperature measured at $\nu=0.21$ at $22$ kHz and $1$ kHz.}
  \label{fig:newfigS1}
\end{figure*}

\begin{figure}[tbh!]
  \begin{center}
    \psfig{file=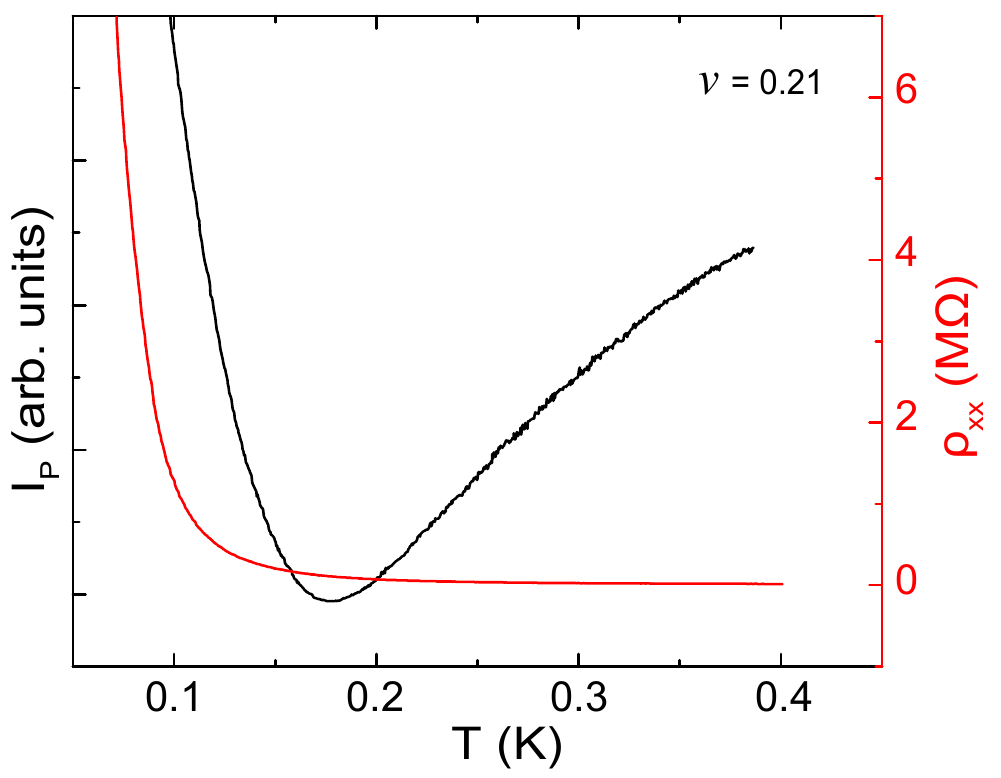, width=0.48\textwidth}
  \end{center}
  \caption{\label{newfigS2}
  The penetrating current $I_P$ and the longitudinal resistivity $\rho_{xx}$ vs. temperature at filling factor $\nu=0.21$ for sample A at $p=3.8 \times 10^{10}$ cm$^{-2}$.}
  \label{fig:newfigS2}
\end{figure}

Figure \ref{fig:newfigS1} shows a summary of the frequency dependence measurements for sample A at $p=3.8 \times 10^{10}$ cm$^{-2}$. Figure \ref{fig:newfigS1}(a) shows the penetrating current $I_P$ vs. magnetic field $B$ at $22$ and $1$ kHz respectively. Despite the quantitative difference in the magnitude of $I_P$, the traces show qualitatively the same behavior. Figure \ref{fig:newfigS1}(b) shows the temperature dependence of $I_P$ at $22$ and $1$ kHz, measured at $\nu=0.21$. The critical temperature $T_C$ at which $I_P$ shows a minimum is the same even though the measurement frequencies differ by more than an order of magnitude, indicating that the measured critical temperature is independent of the measurement frequency.

Figure \ref{fig:newfigS2} compares the temperature dependence of the screening efficiency ($I_P$) and the transport ($\rho_{xx}$) measurements at $\nu=0.21$ for sample A at $p=3.8 \times 10^{10}$ cm$^{-2}$. In sharp contrast to the monotonic decrease of $\rho_{xx}$ as the temperature is increased, $I_P$ decreases first and then increases, showing a well-defined minimum at a critical temperature $T_C$. A qualitatively similar behavior was reported in Ref. \cite{Deng.PRL.2019} for the magnetic-field-induced WS states near $\nu=1/5$ in GaAs 2D electrons. The contrast between the $I_P$ and $\rho_{xx}$ traces implies that the $I_P$ measurements provide additional information which is not discernible in transport measurements. While we do not have a clear explanation for why this is so, we speculate that the $I_P$ measurements might be more sensitive to the presence of a possible intermediate phase \cite{Knighton.PRB.2018} or the additional dissipation from mobile dislocations and uncondensed charge carriers near the WS melting temperature \cite{Delacretaz.PRB.2019}, as we mentioned in the main text.

\section{Measurement results at additional densities}

\subsection{Data for $p=6.2 \times 10^{10}$ cm$^{-2}$}

\begin{figure}[b!]
  \begin{center}
    \psfig{file=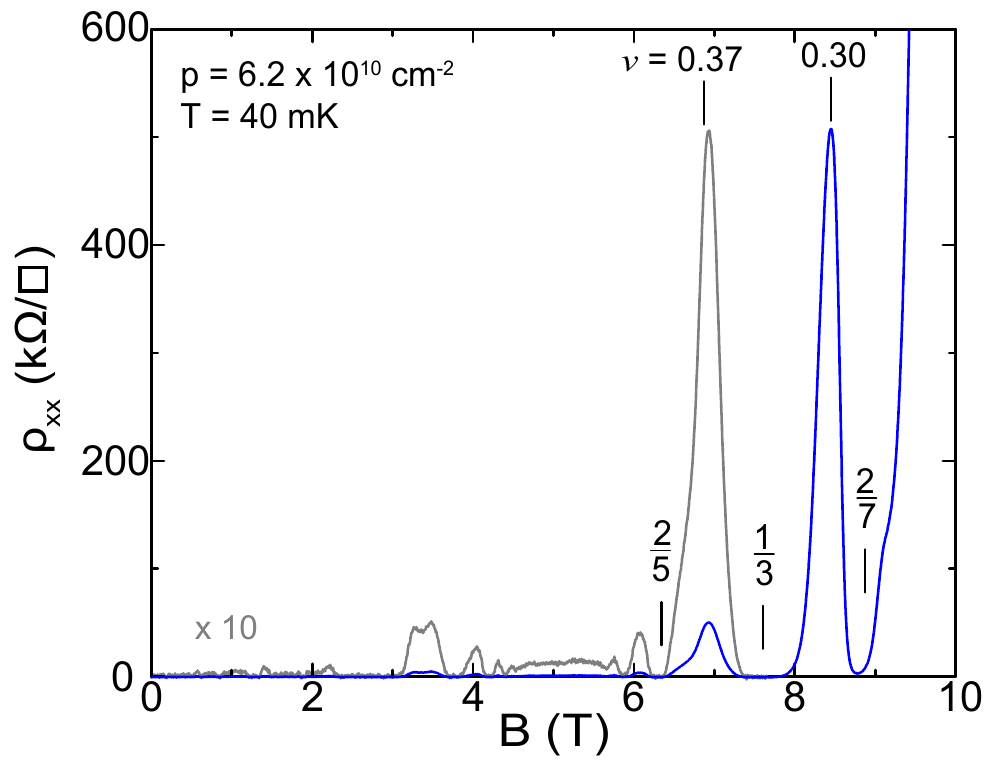, width=0.46\textwidth}
  \end{center}
  \caption{\label{figS14}
  The longitudinal resistivity $\rho_{xx}$ vs. perpendicular magnetic field $B$ for sample A at a density of $p=6.2 \times 10^{10}$ cm$^{-2}$. The traces are measured at a base temperature of $\simeq 40$ mK. The low-field part of the trace ($B<7.5$ T) is shown in grey, with its $y$-scale amplified by a factor of 10. The vertical marks indicate the expected positions of the labeled filling factors.}
  \label{fig:FigS14}
\end{figure}

Figure \ref{fig:FigS14} shows the longitudinal resistivity $\rho_{xx}$ vs. perpendicular magnetic field $B$ for sample A at $p=6.2 \times 10^{10}$ cm$^{-2}$. The trace was taken at a base temperature of $\simeq 40$ mK. The vertical marks indicate the expected positions of the filling factors as labeled. At this density, the $\nu=1/3$, $2/5$, and $2/7$ fractional quantum Hall states (FQHSs) are fully developed but the values of $\rho_{xx}$ at fillings between the FQHSs, namely at $\nu=0.37$ and $0.30$, are quite high, consistent with disorder-pinned WS phases. In particular, $\rho_{xx}$ at $\nu=0.30$ is $\simeq 507$ k$\Omega/\square$. At $\nu=0.37$ $\rho_{xx}$ is $\simeq 50$ k$\Omega/\square$, about an order of magnitude larger than the $\rho_{xx}$ peaks at lower magnetic fields. When we raise the current from $1$ nA to $5$ nA, $\rho_{xx}$ at $\nu=0.37$ drops by a factor of two, but stays nearly constant at lower fields. We therefore conclude that at this density, the 2DHS is likely to be a pinned WS at $\nu=0.37$ and $\nu=0.30$.

\subsection{Data for $p=3.7 \times 10^{10}$ cm$^{-2}$}

\begin{figure}[b!]
  \begin{center}
    \psfig{file=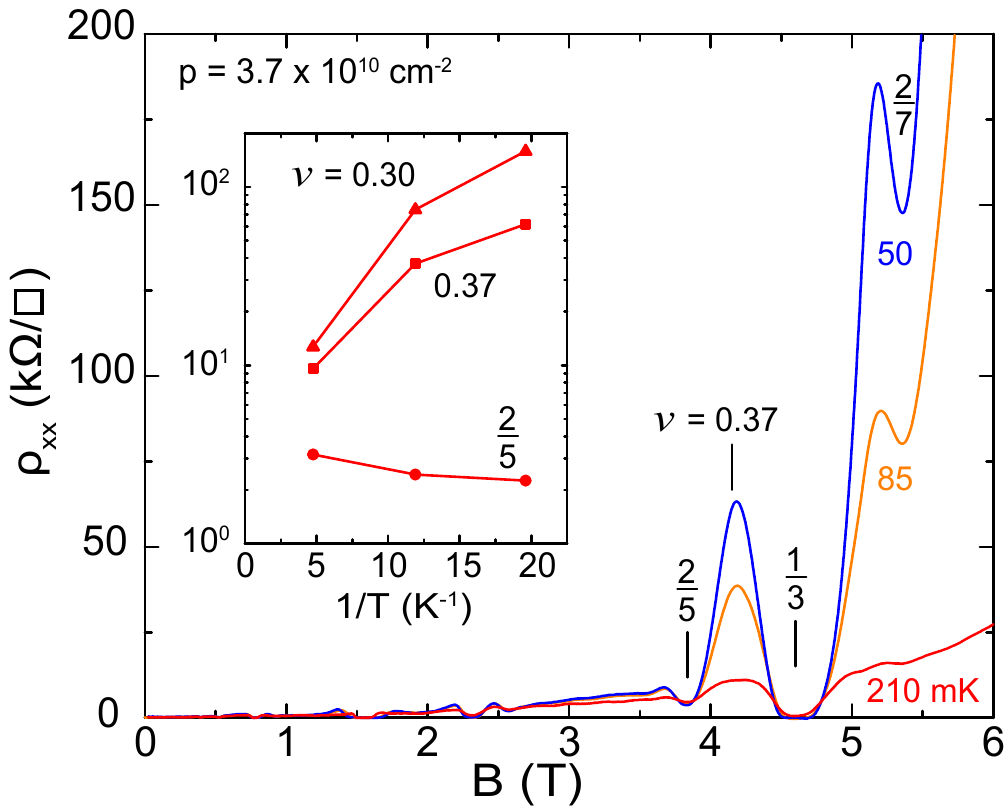, width=0.46\textwidth}
  \end{center}
  \caption{\label{figS1} 
  Temperature dependence of $\rho_{xx}$ vs. $B$ for sample C at a density of $p=3.7 \times 10^{10}$ cm$^{-2}$ with both the front and back gates grounded. The three representative traces are taken at $50$, $85$, and $210$ mK as indicated. The inset shows the Arrhenius plots of $\rho_{xx}$ at filling factors $\nu=0.30$, $0.37$ and $2/5$.}
\label{fig:FigS1}
\end{figure}

\begin{figure}[t!]
  \begin{center}
    \psfig{file=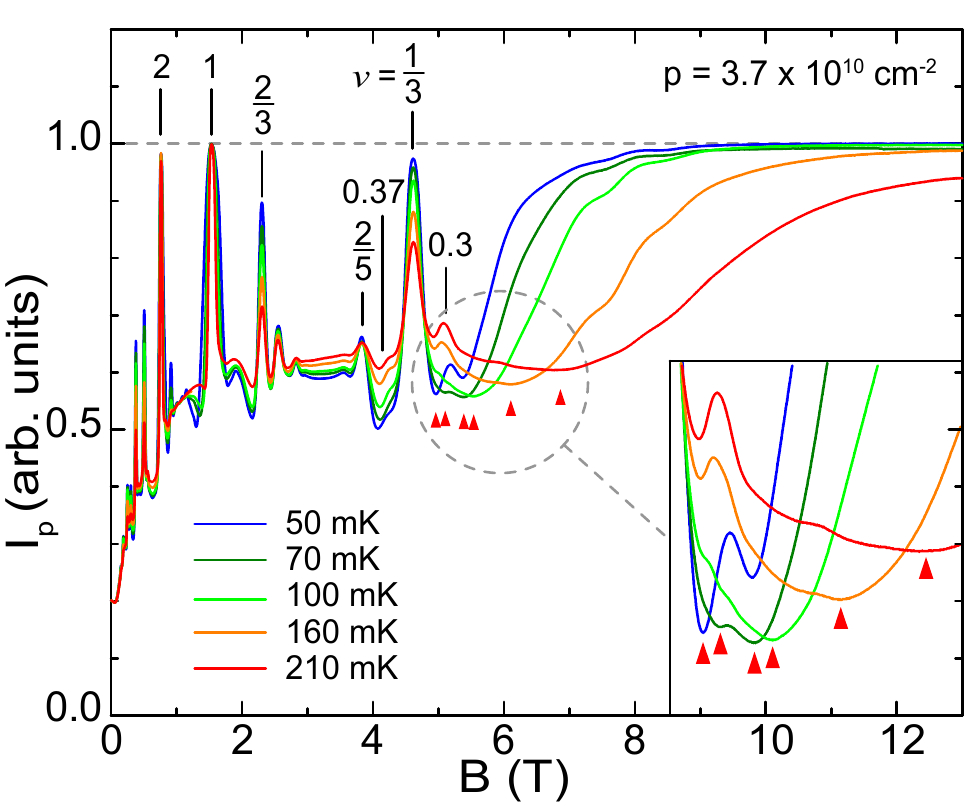, width=0.456\textwidth}
  \end{center}
  \caption{\label{figS2}
  The penetrating current $I_P$, normalized to its maximum value, vs. magnetic field $B$ measured at various temperatures for sample C at $p=3.7 \times 10^{10}$ cm$^{-2}$. The horizontal dashed line shows the maximum $I_P$ when the 2DHS screening is minimal. The vertical lines mark the field positions of fillings $\nu=2$, $1$, $2/3$, $2/5$, $0.37$, $1/3$ and $0.30$, where $I_P$ shows local maxima. The red triangles indicate the $I_P$ minima positions at each temperature and they correspond to the black open squares in Fig. \ref{fig:FigS3} and the blue open circles in Fig. \ref{fig:FigS12}.}
  \label{fig:FigS2}
\end{figure}

\begin{figure}[t!]
  \begin{center}
    \psfig{file=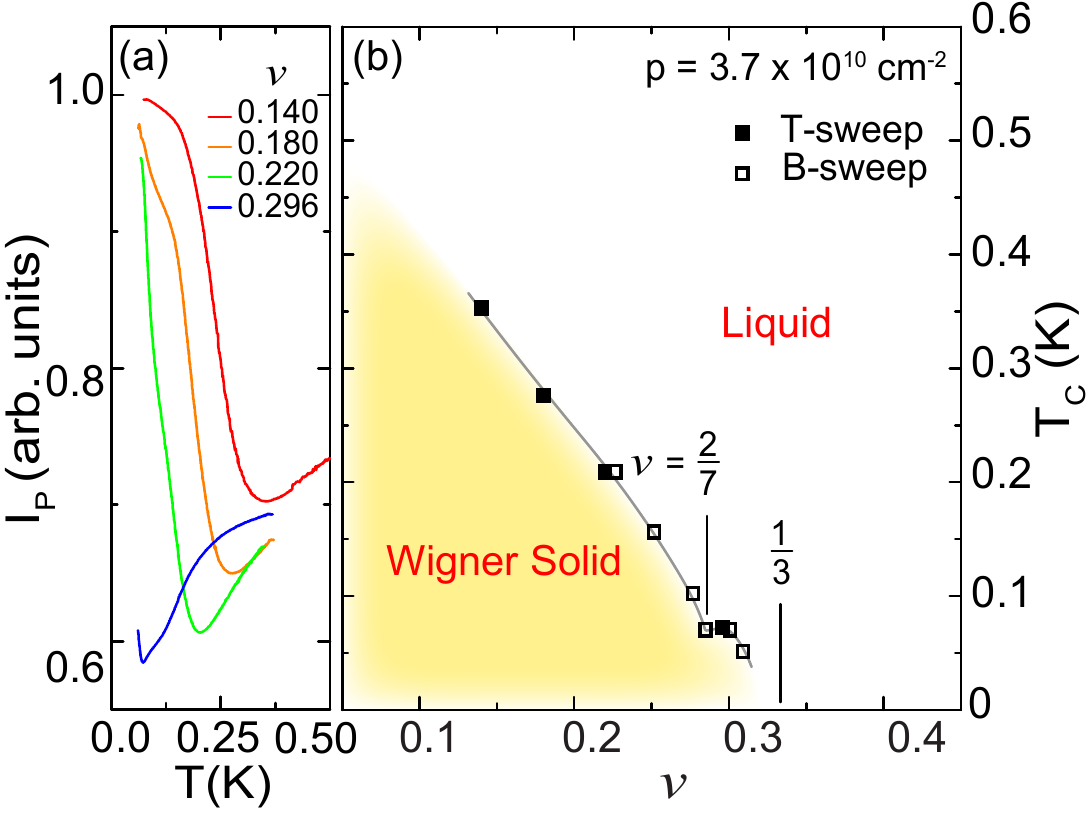, width=0.49\textwidth}
  \end{center}
  \caption{\label{figS3} 
  (a) The penetrating current $I_P$ vs. temperature $T$ for filling factor $\nu$ from $0.140$ to $0.296$ for sample C at $p=3.7 \times 10^{10}$ cm$^{-2}$. (b) The WS critical temperature $T_C$ vs. $\nu$ phase diagram. The yellow region indicates the solid phase and the white region the liquid phase. The black solid squares are temperatures at which $I_P$ show minima for various $\nu$ extracted from (a), and the black open squares correspond to the positions of the red triangles in Fig. \ref{fig:FigS2}. The vertical marks indicate the expected positions of the fillings $\nu=1/3$ and $2/7$. The grey line is a guide to the eye.}
  \label{fig:FigS3}
\end{figure}

In this section we present additional data for sample C at $p=3.7 \times 10^{10}$ cm$^{-2}$. Figure \ref{fig:FigS1} shows the temperature dependence of $\rho_{xx}$ vs. $B$, measured at $50$, $85$, and $210$ mK. Similar to the data measured at $p=3.8 \times 10^{10}$ cm$^{-2}$, shown in Fig. 2 of the main text, the $\nu=1/3$ FQHS is fully developed at the lowest temperature while on its flanks $\rho_{xx}$ shows high resistivity consistent with an insulating phase. The developing $2/7$ FQHS is also prominent here. The inset shows the Arrhenius plot of $\rho_{xx}$ at filling factors $\nu=0.30$, $0.37$ and $2/5$. The $\rho_{xx}$ values are comparable to those of Fig. 2 in the main text at similar temperatures; also, $\rho_{xx}$ at $\nu=0.30$ and $0.37$ increases with decreasing temperature, while at $\nu=2/5$, it decreases.

Figure \ref{fig:FigS2} shows the  penetrating current $I_P$ vs. $B$ measured at various temperatures. The trace at the lowest temperature of $50$ mK shows features qualitatively similar to those at $p=3.8 \times 10^{10}$ cm$^{-2}$ at $40$ mK, as shown in Fig. 2 of the main text. 

The six red triangles in Fig. \ref{fig:FigS2} indicate the magnetic field positions at which $I_P$ shows local minima. As described in Ref. \cite{Deng.PRL.2019}, this is an alternative way for probing the melting temperature of the WS.

Figure \ref{fig:FigS3} provides a summary of the WS melting phase diagram at this density. Figure \ref{fig:FigS3}(a) shows the temperature sweeps at four filling factors ranging from $\nu=0.140$ to $0.296$. The positions at which $I_P$ shows minimum values are shown by black squares in Fig. \ref{fig:FigS3}(b). The open squares correspond to the magnetic field positions indicated by the red triangles in Fig. \ref{fig:FigS2}. The yellow region indicates the WS phase and the white region the liquid phase. The vertical marks indicate the positions of $\nu=1/3$ and $2/7$. In our measurements on this sample, we could only achieve temperatures down to $\approx 50$ mK, and the WS phase on the higher filling side of $1/3$ is already melted at this temperature: At $\nu=0.37$, we found $I_P$ to be monotonically increasing as a function of temperature for $T\geq 50$ mK.

\subsection{Data for $p=2.9 \times 10^{10}$ cm$^{-2}$}

\begin{figure}[b!]
  \begin{center}
    \psfig{file=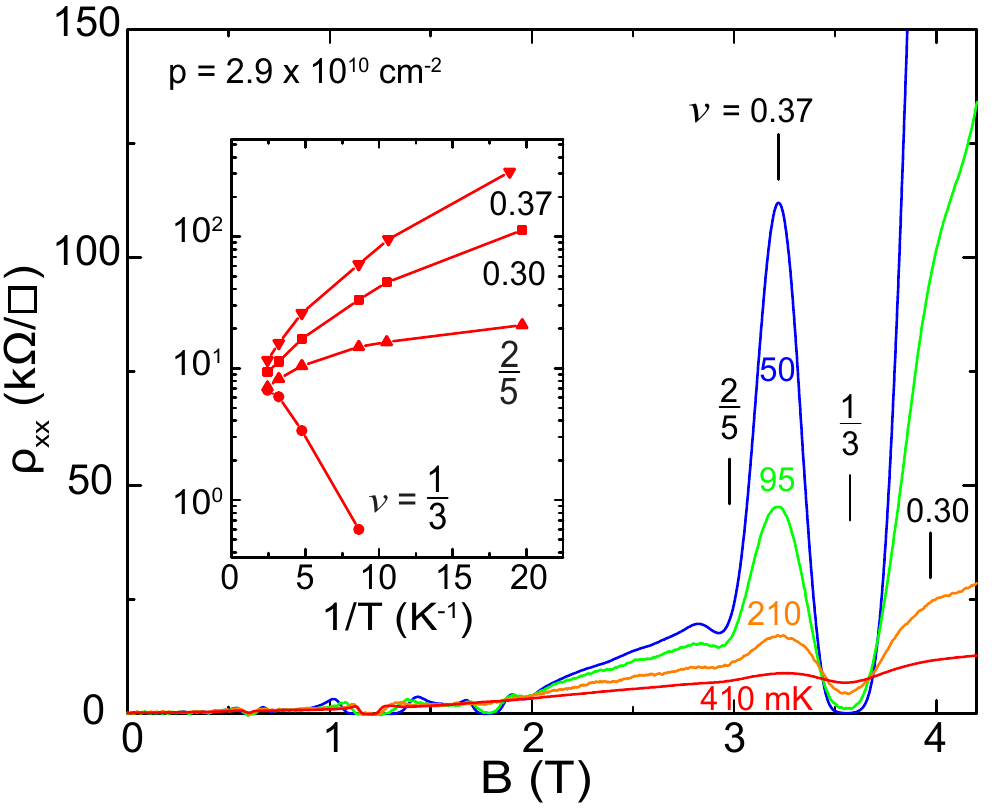, width=0.46\textwidth}
  \end{center}
  \caption{\label{figS4}
  Temperature dependence of $\rho_{xx}$ vs. $B$ for sample C at a density of $p=2.9 \times 10^{10}$ cm$^{-2}$. The four representative traces in the main figure are taken at $50$, $95$, $210$, and $410$ mK. The inset shows the Arrhenius plot of $\rho_{xx}$ at $\nu=0.30$, $0.37$, $2/5$ and $1/3$.}
  \label{fig:FigS4}
\end{figure}

\begin{figure}[t!]
  \begin{center}
    \psfig{file=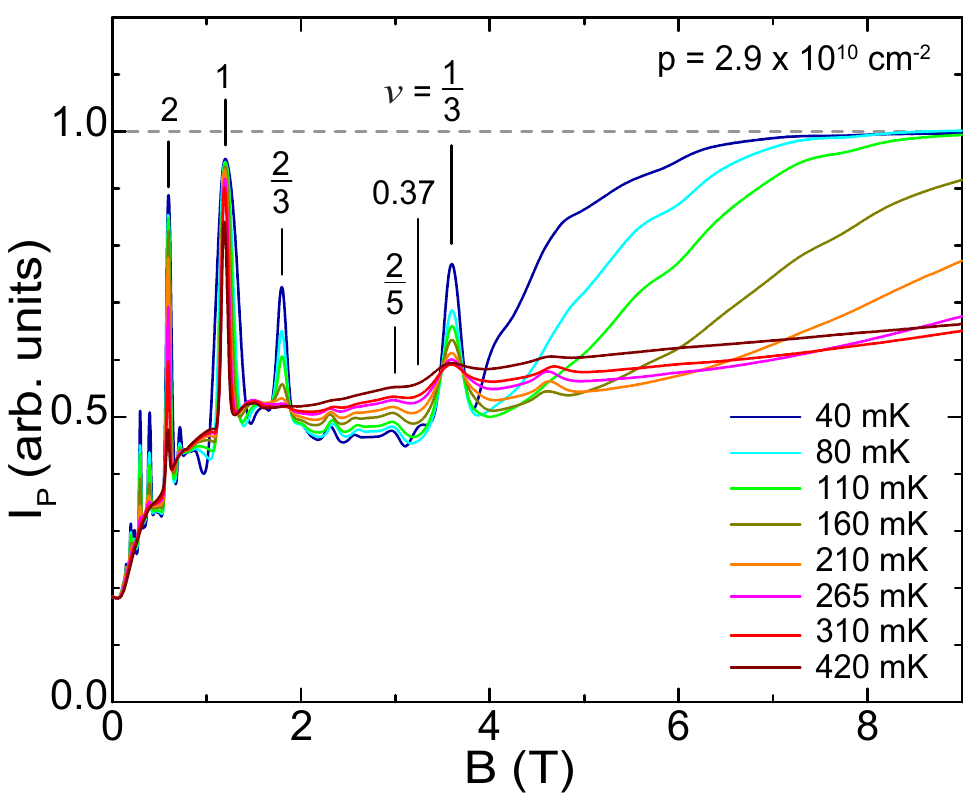, width=0.46\textwidth}
  \end{center}
  \caption{\label{figS5}
  The penetrating current $I_P$, normalized to its maximum value, vs. $B$ measured at various temperatures for sample C at $p=2.9 \times 10^{10}$ cm$^{-2}$. The horizontal dashed line shows the maximum of $I_P$ when the 2DHS screening is minimal. The vertical marks indicate the field positions of fillings $\nu=2$, $1$, $2/3$, $2/5$, $0.37$ and $1/3$, where $I_P$ shows local maxima at the lowest temperatures.}
  \label{fig:FigS5}
\end{figure}

\begin{figure}[b!]
  \begin{center}
    \psfig{file=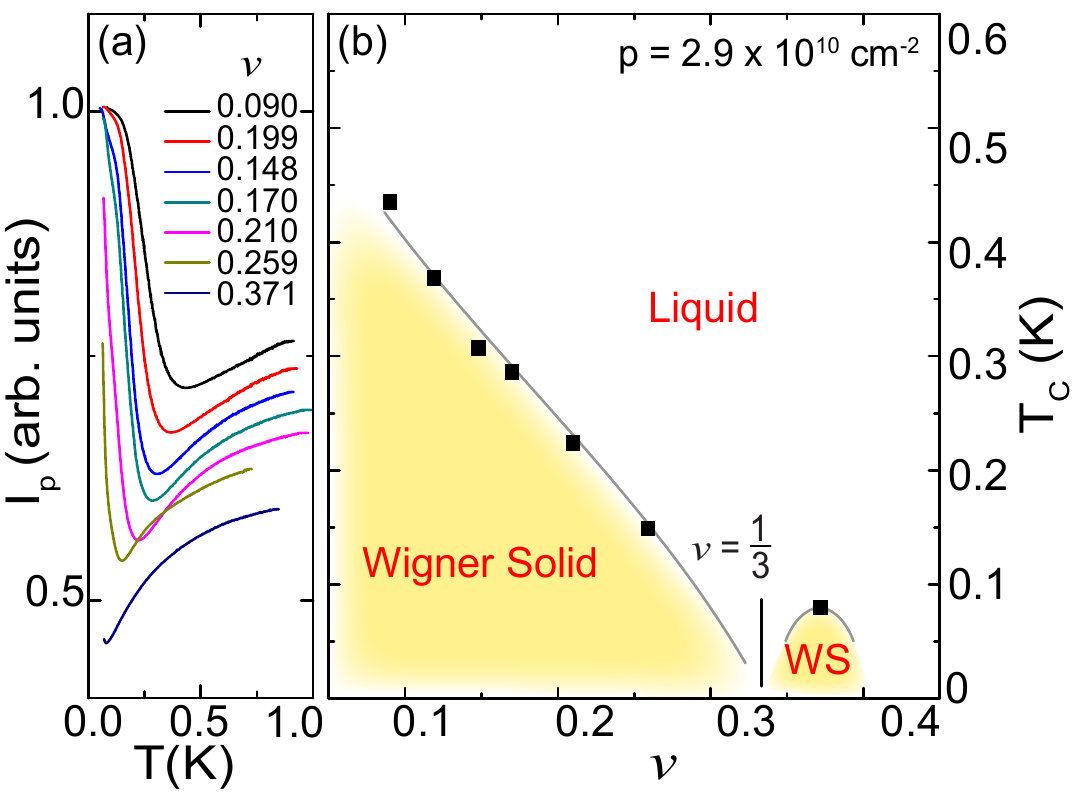, width=0.49\textwidth}
  \end{center}
  \caption{\label{figS6}
  (a) The penetrating current $I_P$ vs. temperature $T$ for filling factor $\nu$ from $0.090$ to $0.371$ for sample C at $p=2.9 \times 10^{10}$ cm$^{-2}$. (b) The WS critical temperature $T_C$ vs. $\nu$ phase diagram. The yellow region indicates the WS phase and the white region the liquid phase. The black solid squares are temperatures at which $I_P$ show minima for various $\nu$ extracted from (a). The vertical line indicates the position of the $\nu=1/3$ FQHS. The grey line is a guide to the eye.}
  \label{fig:FigS6}
\end{figure}

In this section we present additional measurement results for sample C at $p=2.9 \times 10^{10}$ cm$^{-2}$. Figure \ref{fig:FigS4} shows the temperature dependence of $\rho_{xx}$ vs. $B$ measured at $50$, $95$, $210$, and $410$ mK. At this density, the $\nu=1/3$ FQHS is fully developed. There is a sign of an insulating background around $\nu=2/5$, and the $2/7$ FQHS is no longer present. The inset shows the Arrhenius plot of $\rho_{xx}$ at $\nu=0.30$, $0.37$, $2/5$, and $1/3$. $\rho_{xx}$ at $\nu=0.30$, $0.37$, and $2/5$ shows an insulating behavior, in stark contrast to the behavior at $\nu=1/3$, which is activated with a FQHS energy gap of $\simeq0.86$ K.

Figure \ref{fig:FigS5} shows the  $I_P$ vs. $B$ measured at different temperatures. From the lowest temperature trace we can see that the quantum Hall states are all weaker compared to those seen in Fig. \ref{figS2}. Furthermore, the local maximum at $\nu=0.37$ at the lowest temperatures is stronger. At high magnetic fields, $I_P$ starts to rise right after the $1/3$ FQHS to the saturating value, and the local maximum at $\nu=0.30$ is washed out by the rapidly rising background. 

Figure \ref{fig:FigS6} shows the measured WS melting phase diagram at this density. Figure \ref{fig:FigS6}(a) shows the temperature sweeps of $I_P$ at filling factors ranging from $\nu=0.090$ to $0.371$. The critical temperatures at which $I_P$ shows minimum values are shown as black squares in Fig. \ref{fig:FigS6}(b). The yellow region indicates the WS phase and the white region the liquid phase. The vertical line indicates the position of the $\nu=1/3$ FQHS, and as shown here, the WS phase on the higher filling side of $1/3$ is now present above $50$ mK.

\subsection{Data for $p=2.0 \times 10^{10}$ cm$^{-2}$}

\begin{figure}[b!]
  \begin{center}
    \psfig{file=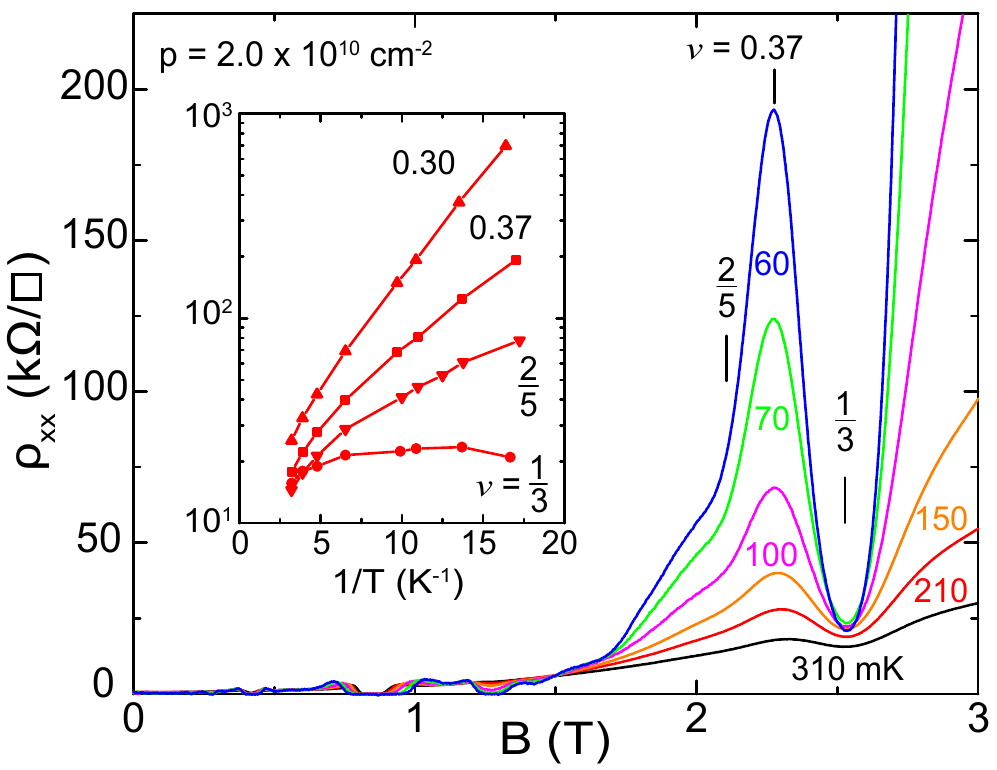, width=0.46\textwidth}
  \end{center}
  \caption{\label{figS15}
  Temperature dependence of $\rho_{xx}$ vs. $B$ for sample C at $p=2.0 \times 10^{10}$ cm$^{-2}$. The six representative traces are taken between $60$ and $310$ mK, as indicated. The vertical marks indicate the field positions of filling factors $\nu=2/5$, $0.37$, and $1/3$. The inset shows the Arrhenius plot of $\rho_{xx}$ for $\nu=0.30$, $0.37$, $2/5$ and $1/3$.}
  \label{fig:FigS15}
\end{figure}

\begin{figure}[t!]
  \begin{center}
    \psfig{file=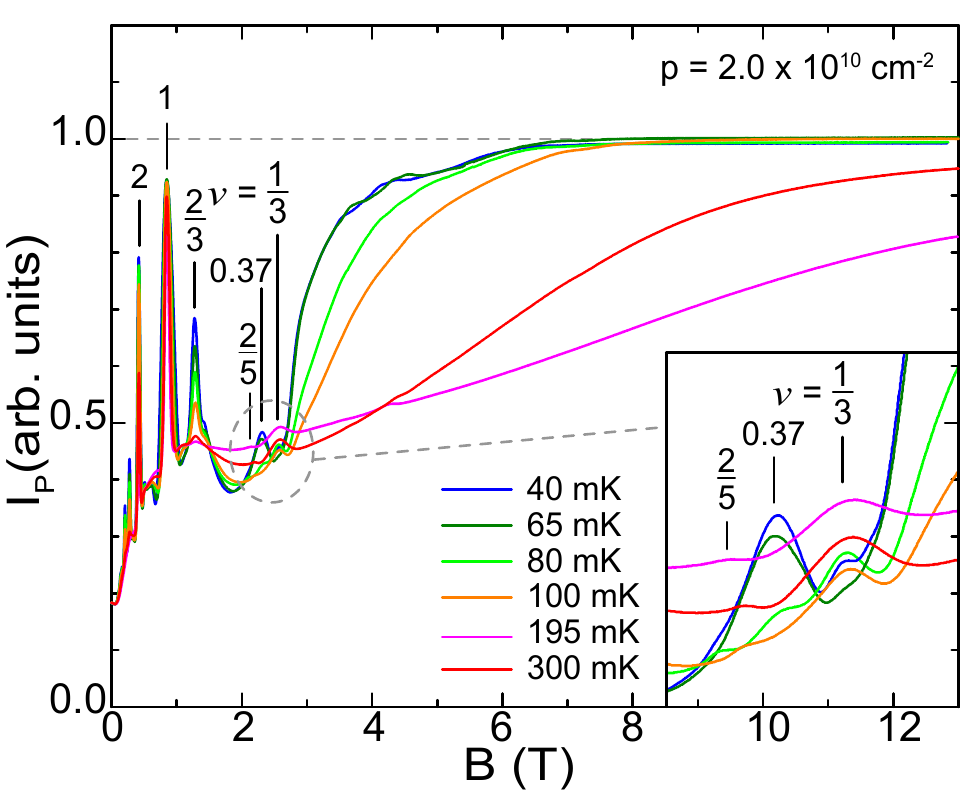, width=0.46\textwidth}
  \end{center}
  \caption{\label{figS17}
  The penetrating current $I_P$, normalized to its maximum value, vs. $B$ measured for sample C at various temperatures at $p=2.0 \times 10^{10}$ cm$^{-2}$. The horizontal dashed line shows the maximum $I_P$ when the 2DHS screening is minimal. The vertical marks indicate the field positions of fillings $\nu=2$, $1$, $2/3$, $2/5$, $0.37$ and $1/3$, where $I_P$ shows local maxima.}
  \label{fig:FigS17}
\end{figure}

\begin{figure}[b!]
  \begin{center}
    \psfig{file=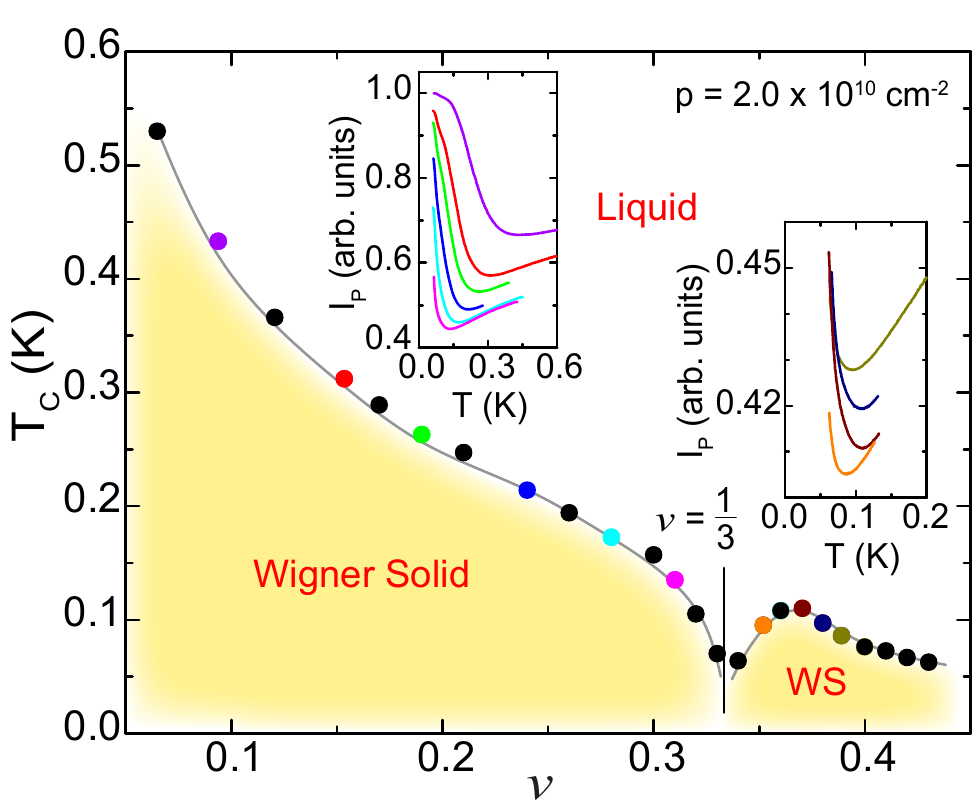, width=0.46\textwidth}
  \end{center}
  \caption{\label{figS16}
  The WS critical temperature $T_C$ vs. $\nu$ phase diagram for sample C at $p=2.0 \times 10^{10}$ cm$^{-2}$. The yellow region indicates the WS phase and the white region the liquid phase. The solid circles are temperatures at which $I_P$ shows minima for various $\nu$; they are color coded to match the $I_P$ vs. $T$ traces in the insets.}
  \label{fig:FigS16}
\end{figure}

Figure \ref{fig:FigS15} shows the temperature dependence of $\rho_{xx}$ vs. $B$ for sample C at $p=2.0 \times 10^{10}$ cm$^{-2}$. There is a lifting of $\rho_{xx}$ background at low temperatures around $\nu=2/5$, which is now so strong that there is no longer a minimum at $\nu=2/5$. At $\nu=1/3$, $\rho_{xx}$ increases with decreasing temperature and starts to decrease only at the lowest temperature, consistent with previous low-density measurements \cite{Csathy.PRL.2004}. The inset shows the Arrhenius plot of $\rho_{xx}$ at $\nu=1/3$, $2/5$, $0.37$, and $0.30$. At this low density the insulating behavior at $\nu=0.30$, $0.37$, and $2/5$ is much stronger, meaning that $\rho_{xx}$ decreases more as temperature rises compared to higher densities, and $\rho_{xx}$ at $\nu=1/3$ becomes non-monotonic with temperature. 

Figure \ref{fig:FigS17} shows the  $I_P$ vs. $B$ measured at various temperatures. At the lowest temperature, the $\nu=1/3$ FQHS is much weaker at this density. On the other hand, the $I_P$ peak corresponding to the insulating phase at $\nu=0.37$ becomes stronger. When the temperature is raised, the peak at $\nu=0.37$ vanishes quickly, indicating the melting of the WS while the peak at $\nu=1/3$ lasts even up to $300$ mK.

Figure \ref{fig:FigS16} shows the deduced melting phase diagram. The two insets show the temperature dependence of $I_P$ at a number of filling factors on the left and right sides of $\nu=1/3$ respectively, color coded to match the data points in the main figure. Similar to Fig. \ref{fig:FigS6}, the WS phase exists on both sides of the $\nu=1/3$ FQHS. On the higher-filling side of $1/3$, the phase boundary exhibits a ``dome" shape whose maximum is at $\nu\simeq0.37$. The dome shape appears to be asymmetric. Also, it goes through $\nu=2/5$ instead of coming down and approaching zero at $\nu=2/5$. We attribute this asymmetry to the developing insulating background near $\nu=2/5$ mentioned earlier. 

\section{Phase diagram summary and discussion of the role of disorder}

\begin{figure}[b!]
  \begin{center}
    \psfig{file=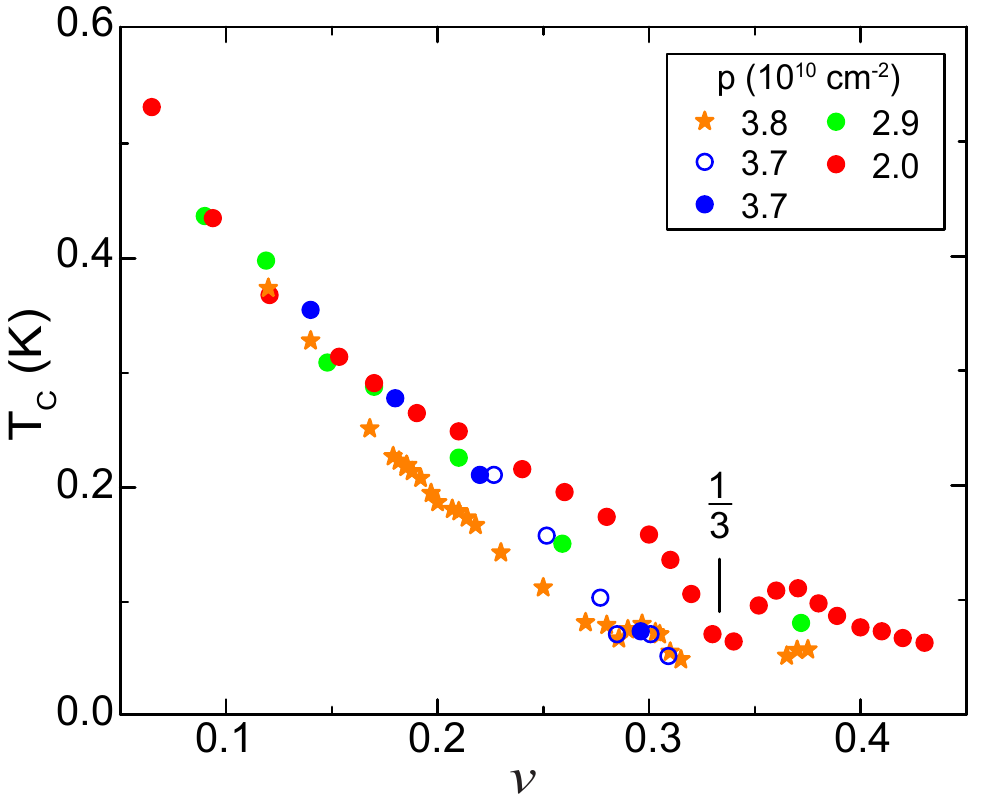, width=0.46\textwidth}
  \end{center}
  \caption{\label{figS12}
  The WS critlcal temperature $T_C$ vs. $\nu$ phase diagram for all measured densities are plotted together for comparison. The stars are data points from sample A and the circles are from sample C. For $p=3.7 \times 10^{10}$ cm$^{-2}$, the closed circles are the measured $T_C$ from temperature sweeps of $I_P$, and the open circles corresponds to the minima positions of $I_P$ vs. $B$ sweep as indicated by the red triangles in Fig. \ref{fig:FigS2}.}
  \label{fig:FigS12}
\end{figure}

Figure \ref{fig:FigS12} shows a summary of the WS melting phase diagrams plotted together for comparison. The data points shown by stars are from sample A and those shown by circles are from sample C. The vertical mark indicates the position of $\nu=1/3$. On the higher filling side of $1/3$, $T_C$ increases as the density decreases. This general trend is also seen on the lower filling side of $1/3$ down to $\nu\sim0.15$. The increase in melting temperature at lower density is unexpected in the clean limit because the Coulomb energy is lower at lower density, and should bring down the melting temperature. This was indeed observed in Ref. \cite{Deng.PRL.2019} for very high quality GaAs two-dimensional electron systems (2DESs). However, microwave resonance studies on 2DESs at very low densities have reported WS melting temperatures that are nearly independent of density \cite{Chen.Nat.Phys.2006}, and have attributed it to the increasing role of the disorder pinning potential at lower densities \cite{Tsukada.JPSJ.1977}. Theoretical work in Ref. \cite{Tsukada.JPSJ.1977} in fact predicts that a stronger disorder (impurity) potential raises the melting temperature of the WS. It is possible that in our 2DHSs, the disorder potential at very low densities plays a strong role, leading to an increase of the melting temperature as the density is lowered. This is also consistent with a previous study of a lower quality 2DHS having a slightly higher melting temperature compared with our data \cite{Bayot.EPL.1994}.


\begin{thebibliography}{99}

\bibitem{Wigner.PR.1934} E. Wigner, On the Interaction of Electrons in Metals, Phys. Rev. {\bf 46}, 1002 (1934).

\bibitem{Lozovik.JETP.Lett.1975} Y. E. Lozovik and V. I. Yudson, Crystallization of a two-dimensional electron gas in a magnetic field, JETP Lett. {\bf 22}, 11 (1975).

\bibitem{Lam.PRB.1984} P. K. Lam and S. M. Girvin, Liquid-solid transition and the fractional quantum-Hall effect, Phys. Rev. B {\bf 30}, 473(R) (1984).

\bibitem{Levesque.PRB.1984} D. Levesque, J. J. Weis, and A. H. MacDonald, Crystallization of the incompressible quantum-fluid state of a two-dimensional electron gas in a strong magnetic field, Phys. Rev. B {\bf 30}, 1056(R) (1984).

\bibitem{Archer.PRL.2013} A. C. Archer, K. Park, and J. K. Jain, Competing Crystal Phases in the Lowest Landau Level, Phys. Rev. Lett {\bf 111}, 146804 (2013).

\bibitem{Tsui.PRL.1982} D. C. Tsui, H. L. Stormer, and A. C. Gossard, Two-Dimensional Magnetotransport in the Extreme Quantum Limit, Phys. Rev. Lett. {\bf 48}, 1559 (1982).

\bibitem{Andrei.PRL.1988} E. Y. Andrei, G. Deville, D. C. Glattli, F. I. B. Williams, E. Paris, and B. Etienne, Observation of a Magnetically Induced Wigner Solid, Phys. Rev. Lett. {\bf 60}, 2765 (1988).

\bibitem{Willett.PRB.1988} R. L. Willett, H. L. Stormer, D. C. Tsui, L. N. Pfeiffer, K. W. West, and K. W. Baldwin, Termination of the series of fractional quantum hall states at small filling factors, Phys. Rev. B {\bf 38}, 7881(R) (1988).

\bibitem{Jiang.PRL.1990} H. W. Jiang, R. L. Willett, H. L. Stormer, D. C. Tsui, L. N. Pfeiffer, and K. W. West, Quantum liquid versus electron solid around $\nu=1/5$ Landau-level filling, Phys. Rev. Lett. {\bf 65}, 633 (1990).

\bibitem{Goldman.PRL.1990} V. J. Goldman, M. Santos, M. Shayegan, and J. E. Cunningham, Evidence for two-dimentional quantum Wigner crystal, Phys. Rev. Lett. {\bf 65}, 2189 (1990).

\bibitem{Williams.PRL.1991} F. I. B. Williams, P. A. Wright, R. G. Clark, E. Y. Andrei, G. Deville, D. C. Glattli, O. Probst, B. Etienne, C. Dorin, C. T. Foxon, and J. J. Harris, Conduction threshold and pinning frequency of magnetically induced Wigner solid, Phys. Rev. Lett. {\bf 66}, 3285 (1991).

\bibitem{Li.PRL.1991} Y. P. Li, T. Sajoto, L. W. Engel, D. C. Tsui, and M. Shayegan, Low-frequency noise in the reentrant insulating phase around the $1/5$ fractional quantum Hall liquid, Phys. Rev. Lett. {\bf 67}, 1630 (1991).

\bibitem{Jiang.PRB.1991} H. W. Jiang, H. L. Stormer, D. C. Tsui, L. N. Pfeiffer, and K. W. West, Magnetotransport studies of the insulating phase around $\nu=1/5$ Landau-level filling, Phys. Rev. B {\bf 44}, 8107 (1991).

\bibitem{Paalanen.PRB.1992} M. A. Paalanen, R. L. Willett, R. R. Ruel, P. B. Littlewood, K. W. West, and L. N. Pfeiffer, Electrical conductivity and Wigner crystallization, Phys. Rev. B(R) {\bf 45}, 13784(R) (1992).

\bibitem{Goldys.PRB.1992} E. M. Goldys, S. A. Brown, R. B. Dunford, A. G. Davies, R. Newbury, R. G. Clarck, P. E. Simmonds, J. J. Harris, and C. T. Foxon, Magneto-optical probe of two-dimensional electron liquid and solid phases, Phys. Rev. B {\bf 46}, 7957(R) (1992).

\bibitem{Kukushkin.Euro.Phys.Lett.1993} I. V. Kukushkin, N. J. Pulsford, K. v. Klitzing, R. J. Haug, K. Ploog, and V. B. Timofeev, Wigner Solid vs. Incompressible Laughlin Liquid: Phase Diagram Derived from Time-Resolved Photoluminescence, Europhys. Lett. {\bf 23}, 211 (1993).

\bibitem{Shayegan.WS.Review.1997} For an early review, see M. Shayegan, Case for the magnetic-field-induced two-dimensional Wiger crystal, in \textit{Perspectives in Quantum Hall Effects}, edited by S. D. Sarma and A. Pinczuk (Wiley, New York, 1997), pp. 343-383.

\bibitem{Pan.PRL.2002} W. Pan, H. L. Stormer, D. C. Tsui, L. N. Pfeiffer, K. W. Baldwin, and K. W. West, Transition from an Electron Solid to the Sequence of Fractional Quantum Hall States at Very Low Landau Level Filling Factor, Phys. Rev. Lett. {\bf 88}, 176802 (2002).

\bibitem{Ye.PRL.2002} P. D. Ye, L. W. Engel, D. C. Tsui, R. M. Lewis, L. N. Pfeiffer, and K. West, Correlation Lengths of the Wigner-Crystal Order in a Two-Dimensional Electron System at High Magnetic Fields, Phys. Rev. Lett. {\bf 89}, 176802 (2002).

\bibitem{Chen.Nat.Phys.2006} Y. P. Chen, G. Sambandamurthy, Z. H. Wang, R. M. Lewis, L. W. Engel, D. C. Tsui, P. D. Ye, L. N. Pfeiffer, and K. W. West, Melting of a 2D quantum electron solid in high magnetic field, Nat. Phys. {\bf 2}, 452 (2006).

\bibitem{Tiemann.Nat.Phys.2014} L. Tiemann, T. D. Rhone, N. Shibata, and K. Muraki, NMR profiling of quantum electron solids in high magnetic fields, Nat. Phys. {\bf 10}, 648 (2014).

\bibitem{Deng.PRL.2016} H. Deng, Y. Liu, I. Jo, L. N. Pfeiffer, K. W. West, K. W. Baldwin, and M. Shayegan, Commensurability Oscillations of Composite Fermions Induced by the Periodic Potential of a Wigner Crystal, Phys. Rev. Lett. {\bf 117}, 096601 (2016).

\bibitem{Jang.Ashoori.Nat.Phys.2017} J. Jang, B. M. Hunt, L. N. Pfeiffer, K. W. West, and R. C. Ashoori, Sharp tunnelling resonance from the vibrations of an electronic Wigner crystal, Nat. Phys. {\bf13}, 340 (2017).

\bibitem{Deng.PRL.2019} H. Deng, L. N. Pfeiffer, K. W. West, K. W. Baldwin, L. W. Engel, and M. Shayegan, Probing the Melting of a Two-Dimensional Quantum Wigner Crystal via its Screening Efficiency, Phys. Rev. Lett. {\bf 122}, 116601 (2019).

\bibitem{Yoon.PRL.1999} J. Yoon, C. C. Li, D. Shahar, D. C. Tsui, and M. Shayegan, Wigner Crystallization and Metal-Insulator Transition of Two-Dimensional Holes in GaAs at $B=0$, Phys. Rev. Lett. {\bf 82}, 1744 (1999).

\bibitem{Manfra.PRL.2007} M. J. Manfra, E. H. Hwang, S. Das Sarma, L. N. Pfeiffer, K. W. West, and A. M. Sergent, Transport and Percolation in a Low-Density High-Mobility Two-Dimensional Hole System, Phys. Rev. Lett. {\bf 99}, 236402 (2007).

\bibitem{Qiu.Gao.PRL.2012} R. L. J. Qiu, X. P. A. Gao, L. N. Pfeiffer, and K. W. West, Connecting the Reentrant Insulating Phase and the Zero-Field Metal-Insulator Transition in a 2D Hole System, Phys. Rev. Lett. {\bf 108}, 106404 (2012).

\bibitem{Santos.PRL.1992} M. B. Santos, Y. W. Suen, M. Shayegan, Y. P. Li, L. W. Engel, and D. C. Tsui, Observation of a reentrant insulating phase near the 1/3 fractional quantum Hall liquid in a two-dimensional hole system, Phys. Rev. Lett. {\bf 68}, 1188 (1992).

\bibitem{Santos.PRB.1992} M. B. Santos, J. Jo, Y. W. Suen, L. W. Engel, and M. Shayegan, Effect of Landau-level mixing on quantum-liquid and solid states of two-dimensional hole systems, Phys. Rev. B {\bf 46}, 13639(R) (1992).

\bibitem{Bayot.EPL.1994} V. Bayot, X. Ying, M. B. Santos, and M. Shayegan, Thermopower in the Re-entrant Insulating Phase of a Two-dimensional Hole system, Europhys. Lett. {\bf25}, 613 (1994).

\bibitem{Li.PRL.1997} C.-C. Li, L. W. Engel, D. Shahar, D. C. Tsui, and M. Shayegan, Microwave Conductivity Resonance of Two-Dimensional Hole System, Phys. Rev. Lett. {\bf79}, 1353 (1997).

\bibitem{Li.PRB.2000} C.-C. Li, J. Yoon, L. W. Engel, D. Shahar, D. C. Tsui, and M. Shayegan, Microwave resonance and weak pinning in two-dimensional hole systems at high magnetic fields, Phys. Rev. B {\bf61}, 10905 (2000).

\bibitem{Csathy.PRL.2004} G. A. Cs\'athy, D. C. Tsui, L. N. Pfeiffer, and K. W. West, Possible Observation of Phase Coexistence of the $\nu=1/3$ Fractional Quantum Hall Liquid and a Solid, Phys. Rev. Lett, {\bf 92}, 256804 (2004).

\bibitem{Csathy.PRL.2005} G. A. Cs\'{a}thy, Hwayong Noh, D. C. Tsui, L. N. Pfeiffer, and K. W. West, Magnetic-Field-Induced Insulating Phases at Large $r_s$, Phys. Rev. Lett. {\bf 94}, 226802 (2005).

\bibitem{Pan.PRB.2005} W. Pan, G. A. Cs\'{a}thy, D. C. Tsui, L. N. Pfeiffer, and K. W. West, Transition from a fractional quantum Hall liquid to an electron solid at Landau level filling $\nu =1/3$ in tilted magnetic fields, Phys. Rev. B {\bf71}, 035302 (2005).

\bibitem{Jo.PRL.2018} I. Jo, H. Deng, Y. Liu, L. N. Pfeiffer, K. W. West, K. W. Baldwin, and M. Shayegan, Cyclotron Orbits of Composite Fermions in the Fractional Quantum Hall Regime, Phys. Rev. Lett. {\bf 120}, 016802 (2018).

\bibitem{Knighton.PRB.2018} T. Knighton, Z. Wu, and J. Huang, A. Serafin, J. S. Xia, L. N. Pfeiffer, and K. W. West, Evidence of two-stage melting of Wigner solids, Phys. Rev. B {\bf 97}, 085135 (2018).

\bibitem{Zhu.SSC.2007} H. Zhu, K. Lai, D. C. Tsui, S. P. Bayrakci, N. P. Ong, M. Manfra, L. Pfeiffer, and K. West, Density and well width dependences of the effective mass of two-dimensional holes in (100) GaAs quantum wells measured using cyclotron resonance at microwave frequencies, Solid State Commun. {\bf 141}, 510 (2007).

\bibitem{Yoshioka.JPSJ.1984} D. Yoshioka, Effect of the Landau Level Mixing on the Ground State of Two-Dimensional Electrons, J. Phys. Soc. Jpn. {\bf53}, 3740 (1984).

\bibitem{Yoshioka.JPSJ.1986} D. Yoshioka, Excitation Energies of the Fractional Quantum Hall Effect, J. Phys. Soc. Jpn. {\bf55}, 885 (1986).

\bibitem{Zhu.PRL.1993} X. Zhu and S. G. Louie, Wigner crystallization in the fractional quantum Hall regime: A variational quantum Monte Carlo study, Phys. Rev. Lett. {\bf70}, 335, (1993).

\bibitem{Price.PRL.1993} R. Price, P. M. Platzman, and S. He, Fractional quantum Hall liquid, Wigner solid phase boundary at finite density and magnetic field, Phys. Rev. Lett. {\bf70}, 339 (1993).

\bibitem{Platzman.PRL.1993} P. M. Platzman and R. Price, Quantum freezing of the fractional quantum Hall liquid, Phys. Rev. Lett. {\bf70}, 3487 (1993).

\bibitem{Ortiz.PRL.1993} G. Ortiz, D. M. Ceperley, and R. M. Martin, New stochastic method for systems with broken time-reversal symmetry: 2D fermions in a magnetic field, Phys. Rev. Lett. {\bf71}, 2777 (1993).

\bibitem{Zhao.PRL.2018} J. Zhao, Y. Zhang, and J. K. Jain, Crystallization in the Fractional Quantum Hall Regime Induced by Landau-Level Mixing, Phys. Rev. Lett. {\bf 121}, 116802 (2018).

\bibitem{Maryenko.Nat.Comm.2018} D. Maryenko, A. McCollam, J. Falson, Y. Kozuka, J. Bruin, U. Zeitler, and M. Kawasaki, Composite fermion liquid to Wigner solid transition in the lowest Landau level of zinc oxide, Nat. Comm. {\bf 9}, 4356 (2018).

\bibitem{supplemental} See Supplemental Material, which includes Refs. \cite{Ando.RMP.1982,Jain.Book.2007,Kernreiter.PRB.2013,Andreani.PRB.1987,Winkler.PRB.2005,Tsukada.JPSJ.1977} for more details on sample structure, additional data and discussions.

\bibitem{Ando.RMP.1982} T. Ando, A. B. Fowler, and F. Stern, Electronic properties of two-dimensional systems, Rev. Mod. Phys. \textbf{54}, 437 (1982).

\bibitem{Jain.Book.2007} J. K. Jain, \textit{Composite Fermions}, (Cambridge University Press, Cambridge, UK, 2007).

\bibitem{Kernreiter.PRB.2013} T.~Kernreiter, M. Governale, R. Winkler, and U. Z\"ulicke, Suppression of Coulomb exchange energy in quasi-two-dimensional hole systems,
 Phys.\ Rev.~B \textbf{88}, 125309 (2013).

\bibitem{Andreani.PRB.1987} L.~C. Andreani, A.~Pasquarello, and F.~Bassani, Hole subbands in strained GaAs--Ga$_{1-x}$A$_x$As quantum wells: Exact solution of the effectve-mass equation, Phys.\ Rev.~B \textbf{36}, 5887 (1987).

\bibitem{Winkler.PRB.2005} R.~Winkler, E. Tutuc, S. J. Papadakis, S. Melinte,
 M. Shayegan, D. Wasserman, and S. A. Lyon, Anomalous Spin Polarization of GaAs Two-Dimensional Hole Systems, Phys.\ Rev.~B \textbf{72}, 195321 (2005).
 
\bibitem{Tsukada.JPSJ.1977} M. Tsukada, Two-Dimensional Crystallization of the Electrons in MOS Structures Induced by Strong Magnetic Field, J. Phys. Soc. Japan {\bf 42}, 391 (1977).


\bibitem{Eisenstein.PRL.1992} J. P. Eisenstein, L. N. Pfeiffer, and K. W. West, Negative Compressibility of Interacting Two-Dimensional Electron and Quasiparticle Gases, Phys. Rev. Lett. {\bf 68}, 674 (1992).

\bibitem{Eisenstein.PRB.1994} J. P. Eisenstein, L. N. Pfeiffer, and K. W. West, Compressibility of the two-dimensional electron gas: Measurements of the zero-field exchange energy and fractional quantum Hall gap, Phys. Rev. B, {\bf 50}, 1760 (1994).

\bibitem{Young.Nat.Phys.2018} A. A. Zibrov, E. M. Spanton, H. Zhou, C. Kometter, T. Taniguchi, K. Watanabe, and A. F. Young, Even-denominator fractional quantum Hall states at an isospin transition in monolayer graphene, Nat. Phys. {\bf 14}, 930 (2018).

\bibitem{Delacretaz.PRB.2019} L. V. Delacr\'{e}taz, B. Goutéraux, S. A. Hartnoll, and A. Karlsson, Theory of collective magnetophonon resonance and melting of a field-induced Wigner solid, Phys. Rev. B {\bf 100}, 085140 (2019).

\bibitem{Winkler.Book.2003} R. Winkler, \textit{Spin-Orbit Coupling Effects in Two-Dimensional Electron and Hole Systems}, Springer Tracts in Modern Physics Vol. 191, (Springer-Verlag, Berlin, 2003).

\end{thebibliography}
\end{document}